\documentclass[11pt,a4paper]{article}
\pdfoutput=1
\usepackage{jheppub,yfonts}

\newcommand{\be}{\begin{equation}}
\newcommand{\ee}{\end{equation}}
\newcommand{\bea}{\begin{eqnarray}}
\newcommand{\eea}{\end{eqnarray}}

\title{Estimates for the Thermal Width of Heavy Quarkonia in Strongly Coupled Plasmas from Holography}

\author{Stefano I.~Finazzo}
\author{and Jorge Noronha}
\affiliation{Instituto de F\'{i}sica, Universidade de S\~{a}o Paulo, S\~{a}o Paulo, SP, Brazil}

\date{\today}

\abstract{The gauge/gravity duality is used to investigate the imaginary part of the heavy quark potential (defined via the rectangular Wilson loop) in strongly coupled plasmas. This quantity can be used to estimate the width of heavy quarkonia in a plasma at strong coupling. In this paper the thermal worldsheet fluctuation method, proposed in [J.~Noronha and A.~Dumitru, Phys.\ Rev.\ Lett.\ {\bf 103}, 152304 (2009)], is revisited and general conditions for the existence of an imaginary part for the heavy quark potential computed within classical gravity models are obtained. We prove a general result that establishes the connection between this imaginary part of the potential determined holographically and the area law displayed by the Wilson loop in the vacuum of confining gauge theories. We also determine the imaginary part of the heavy quark potential in a strongly coupled plasma dual to Gauss-Bonnet gravity. This provides an estimate of how the thermal width of heavy quarkonia changes with the shear viscosity to entropy density ratio, $\eta/s$, at strong coupling.}
   

\emailAdd{noronha@if.usp.br, stefanofinazzo@gmail.com}
 
\begin{document}

\begin{flushright}
\end{flushright}

\maketitle
\setlength{\parskip}{8pt}


\section{Introduction}

\begin{figure}[h!t]
\begin{center}
\includegraphics[width=0.5 \textwidth]{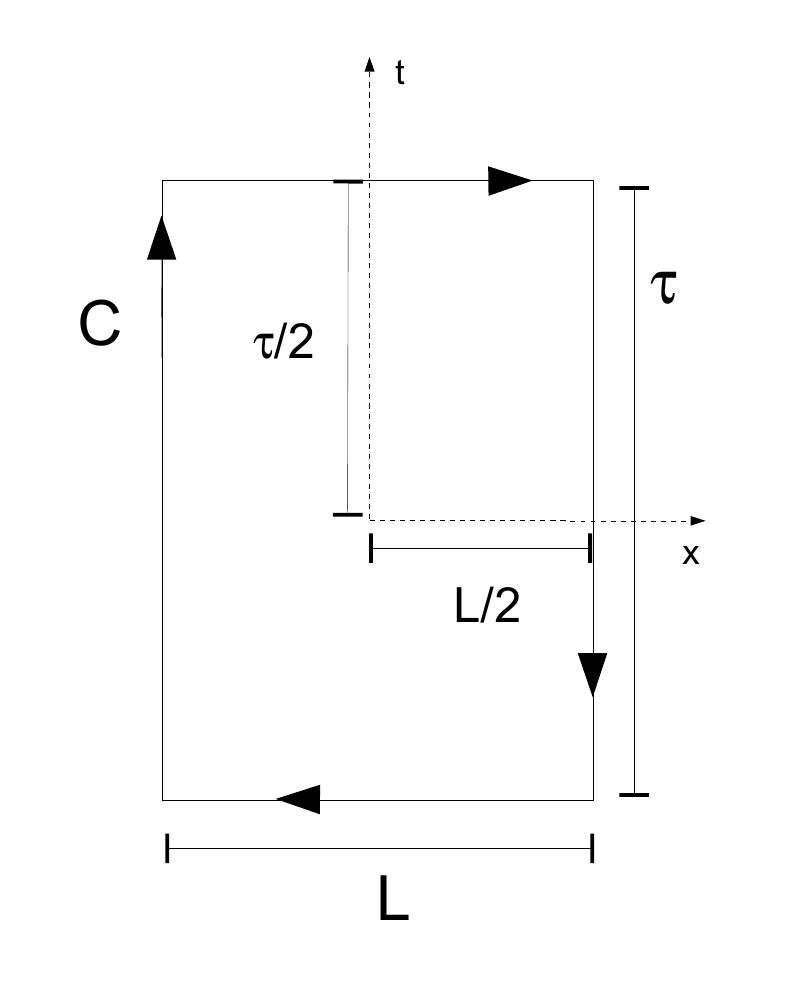}
\caption{The rectangular Wilson loop, along with the choice of the coordinate system used.}
\label{fig:wilsonloop}
\end{center}
\end{figure}

One of the most important gauge invariant quantities defined in non-Abelian $SU(N_c)$ gauge theories \cite{Wilson:1974sk,wilsonloop} is the Wilson loop
\begin{equation}
\label{eq:wilsonloop}
W(C) = \frac{1}{N_c}\textrm{tr} \, P \, \exp{ \left[ i g\oint_C \hat{A}_{\mu} dx^{\mu} \right]},
\end{equation}
where $C$ is a closed loop embedded in a 4-dimensional spacetime, $P$ indicates path-ordering, $g$ is the coupling, $\hat{A}_{\mu}$ is the non-Abelian gauge field potential operator while the trace is performed over the fundamental representation of $SU(N_c)$ (other representations can also be used but we will use the fundamental representation in this paper). In particular, the case where $C$ is a rectangular loop of spatial length $L$ and extended over $\mathcal{T}$ in the time direction, 
as depicted in Fig.\ \ref{fig:wilsonloop}, has been extensively studied over the years. With this contour, the limit $\mathcal{T} \rightarrow \infty$ of the vacuum expectation value of \eqref{eq:wilsonloop} gives
\begin{equation}
\label{eq:wilsonrec}
\lim_{\mathcal{T} \to \infty}\langle W(C) \rangle_0 \sim e^{i \mathcal{T} V_{Q\bar{Q}}(L)},
\end{equation}
where $V_{Q\bar{Q}}(L)$ is known as the heavy quark potential (the vacuum interaction energy between two infinitely massive probes in the fundamental representation). In the vacuum of a confining gauge theory $\langle W(C)\rangle$ should obey an area law defined by $\lim_{L \to \infty}V_{Q\bar{Q}}(L)/L=\sigma$ with $\sigma$ being the string tension \cite{Wilson:1974sk}. \\

In the imaginary time formulation of thermal gauge theories \cite{kapusta}, all bosonic fields are required to be periodic (or anti-periodic in the case of fermionic fields) in the Euclidean time $\tau$ with period $\beta=1/T$ and the order parameter for the deconfinement phase transition in an $SU(N_c)$ theory without dynamical fermions is characterized by the path ordered Polyakov loop \cite{polyakov} 
\begin{equation}
\ \bold{L} (\vec{x})=\frac{1}{N_c}\,P \,e^{i\,g\int_{0}^{1/T} \hat{A}_{0}(\vec{x},\tau)d\tau}\,.
\label{polyakovdef}
\end{equation}
This operator becomes gauge invariant (up to a phase) after performing the trace. In a pure gauge theory there are also global gauge transformations that are only periodic up to an element of $Z(N_c)$, which is the center of $SU(N_c)$. In this case, ${\rm tr\,}\bold{L}$ transforms as a field of charge one under the global $Z(N_c)$ symmetry, i.e., ${\rm tr\,}\bold{L} \to e^{i 2\pi a/N_c}{\rm tr\,}\bold{L}$ where $a=0,\ldots,N_c-1$. Below $T_c$ the system is $Z(N_c)$ symmetric, which implies that $\langle {\rm tr\,}\bold{L} \rangle = 0$. Above $T_c$ this global symmetry is spontaneously broken, $\langle {\rm tr\,}\bold{L} \rangle \neq 0$, and the system lands in one of the possible $Z(N_c)$ vacua. The thermal average of the Polyakov loop correlator $\mathcal{C}(r,T)\equiv\langle {\rm tr}\,\bold{L}^{\dagger}(r)\,{\rm tr}\,\bold{L}(0)\rangle$ is associated with the difference in the free energy of the system due to the inclusion of an infinitely heavy $Q\bar{Q}$ pair separated by a distance $r$ in the medium \cite{McLerran:1980pk}. Such a formulation has been used to define a heavy quark potential at finite temperature on the lattice \cite{Kaczmarek:2002mc,Philipsen:2010gj}. 

However, the rectangular Wilson loop can also be computed in gauge theories at finite temperature. In this case, the expectation value of the Wilson loop operator for the same rectangular contour can be evaluated in a thermal state of the gauge theory with temperature $T$ (in Minkowski spacetime) and the $\mathcal{T} \rightarrow \infty$ limit
\begin{equation}
\label{eq:wilsonrec}
\lim_{\mathcal{T} \to \infty}\langle W(C) \rangle \sim e^{i \mathcal{T} V_{Q\bar{Q}}(L,T)}
\end{equation}   
defines a quantity $V_{Q\bar{Q}}(L,T)$ which we call here the ``heavy quark potential at finite temperature". In general, this heavy quark potential in QCD can have an imaginary part, as shown in \cite{imvrefs,otherrefs1,otherrefsImV,Rothkopf:2011db}, while the quantity defined using the Polyakov loop correlator is necessarily real. The imaginary part of the potential defines a thermal decay width which, at weak coupling, is related to the imaginary part of the gluon self energy induced by Landau damping and the $Q\bar{Q}$ color singlet to color octet thermal break up. 

In this paper we shall elaborate on the method proposed in \cite{Noronha:2009da} to estimate the thermal width of heavy quarkonia at strong coupling using worldsheet fluctuations of the Nambu-Goto action associated with the heavy quark pair in the gauge/gravity duality \cite{gaugegravityduality}. In this approach, the thermal width of heavy quarkonium states  stems from the effect of thermal fluctuations due to the interactions between the heavy quarks and the strongly coupled medium. This is described holographically by integrating out thermal long wavelength fluctuations in the path integral of the Nambu-Goto action in the curved background spacetime. At sufficiently strong coupling, this calculation can be done analytically and a simple formula for the imaginary part of the Wilson loop can be found in this approach that is valid for any gauge theory that is holographically dual to classical gravity \footnote{The background metric has to fulfill certain conditions for the method to be applicable. This is shown in Section \ref{generalformulasection}.}. The formula is used to revisit the calculation of the thermal width in strongly coupled $\mathcal{N}=4$ Super Yang-Mills (SYM) theory done in \cite{Noronha:2009da}. Moreover, we compute the imaginary part of the potential for a strongly-coupled conformal field theory (CFT) dual to Gauss-Bonnet (GB) gravity. We also prove a general result that establishes the connection between the thermal width and the presence of an area law for the Wilson loop at zero temperature in gauge theories with gravity duals, which may be useful for the study of the imaginary part of the heavy quark potential in confining gauge theories dual to gravity.

This paper is organized as follows. In the next section we will revisit the general setup concerning the holographic calculation of Wilson loops. In section \ref{realpartsection} we discuss the holographic calculation of $\mathrm{Re}\,V_{Q\bar{Q}}$, which is necessary to derive our main formula for the imaginary part of the potential in Section \ref{impartsection}. In Section \ref{models} we apply the formula to compute the imaginary part in two different strongly coupled gauge theories with gravity duals. We finish with our conclusions and outlook in Section \ref{conclusions}\footnote{Other aspects of the calculations are presented in Appendices A to D.}. 

\section{Holographic setup}\label{generalformulasection}

After the original calculation of the rectangular Wilson loop in the vacuum of strongly coupled $\mathcal{N}=4$ SYM theory \footnote{Note that in $\mathcal{N}=4$ SYM the Wilson loop operator also contains the 6 adjoint scalars.} by Maldacena \cite{Maldacena:1998im} and its generalization to finite temperature in \cite{Brandhuber:1998bs,Rey:1998bq}, rectangular Wilson loops have been extensively studied in strongly coupled gauge theories using the gauge/gravity duality.   

According to the gauge/gravity prescription \cite{Maldacena:1998im}, the expectation value of $W(C)$ in a strongly coupled gauge theory dual to a theory of gravity is
\begin{equation}
\label{eq:wilsongaugegravity}
\langle W(C) \rangle \sim Z_{str},
\end{equation}
where $Z_{str}$ is the generating functional of the string in the bulk which has the loop $C$ at the boundary. In the 
classical gravity approximation
\begin{equation}
\label{eq:classicalpartition}
Z_{str} \sim e^{i S_{str}},
\end{equation}
where $S_{str}$ is the classical string action propagating in the bulk evaluated at an extremum, $\delta S_{str} = 0$. In the case of a rectangular Wilson loop at nonzero $T$ other extrema can become relevant as one increases the value of $LT$ \cite{Bak:2007fk}. In this paper we are only interested in deeply bound states where $LT <1$ and this question becomes less important\footnote{We shall come back to this point when discussing the calculation of the imaginary part later in Section 5 and also in Appendix D.}. In the classical approximation the worldsheet action $S_{str}$ may be taken as the Nambu-Goto action\footnote{For gravity duals derived within string theory, supersymmetry requires the presence of fermions on the worldsheet but those only enter as an $\hbar$ correction to the action and can be neglected in the supergravity limit in which $\alpha' \to 0$.}
\begin{equation}
\label{eq:nambugoto}
S_{str} = S_{NG} = \frac{1}{2\pi \alpha'} \int d\sigma d\tau \sqrt{-det(G_{\mu\nu} \partial_a X^\mu \partial_b X^\nu}),
\end{equation}
where $X^\mu(\tau,\sigma)$ are the worldsheet embedding coordinates, $\mu,\nu=0,1,\ldots, 4$, $a,b = \sigma, \tau$, and $\alpha'= l_s^2$, where $l_s$ is the string length.


\begin{figure}[h!t]
\begin{center}
\includegraphics[width=1.0 \textwidth]{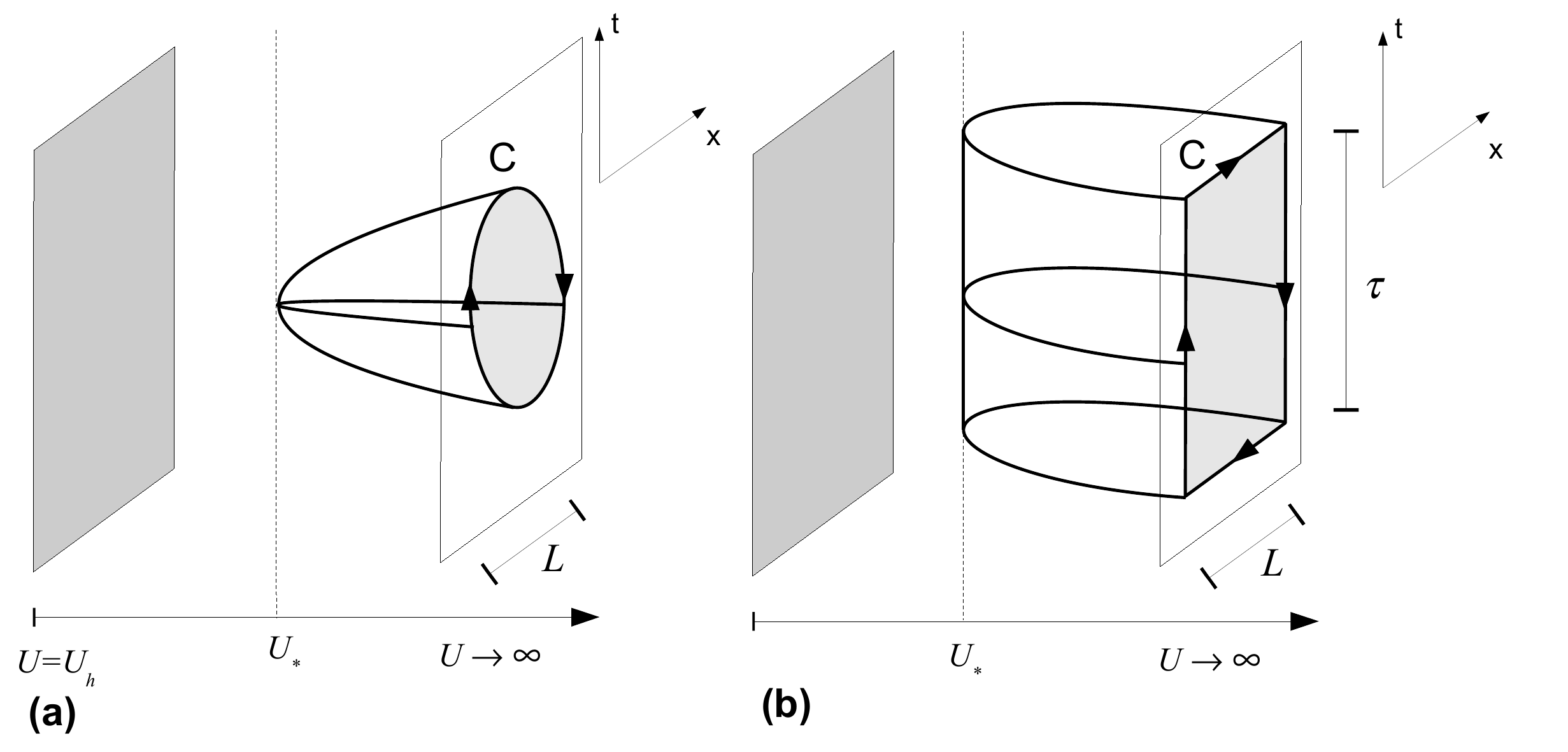}
\caption{The Maldacena prescription for the calculation of Wilson loops via the gauge/gravity duality. In (a) we present the situation for an arbitrary loop $C$. In (b) we consider rectangular Wilson loops with $\mathcal{T} \to \infty$. In both cases $U_h$ is the position of the horizon of the black brane and $U_*$ denotes the bottom of the sagging string in the bulk.}
\label{fig:loopads}
\end{center}
\end{figure}

\begin{figure}[h!t]
\begin{center}
\includegraphics[width=0.5 \textwidth]{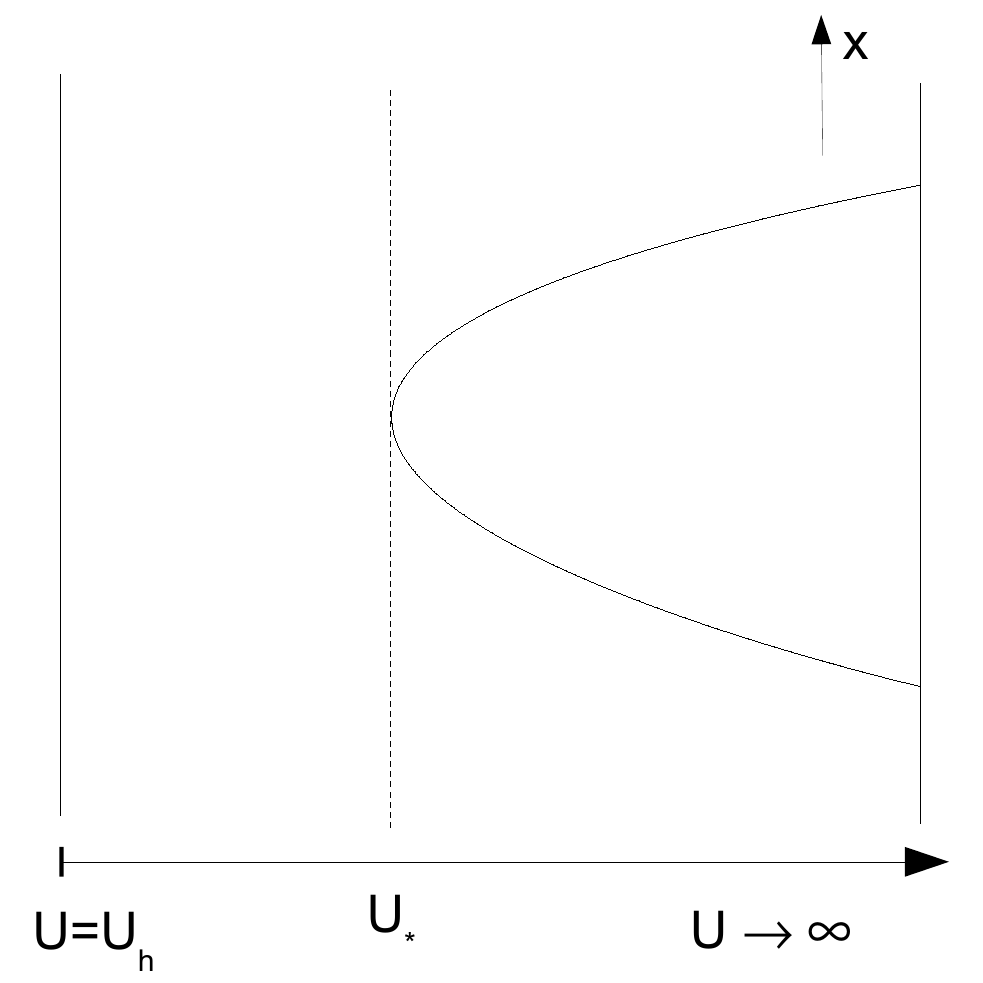}
\caption{A slice of the string worldsheet for the rectangular Wilson loop at fixed time $t$.}
\label{fig:loopadscorte}
\end{center}
\end{figure}

Therefore, the Wilson loop in the strongly coupled gauge theory can be determined using the classical solution of \eqref{eq:nambugoto} which has the loop $C$ as the boundary of the classical string worldsheet. For the case of rectangular Wilson loops one can then calculate $V_{Q\bar{Q}}(L,T)$ (see Fig.\ \ref{fig:loopads}). We will consider an effective 5-dimensional curved spacetime, which will describe the gravity model dual to the gauge theory\footnote{In the case of $AdS\times S_5$ we choose a fixed configuration in $S_5$ for the compact string coordinates.}. Finite temperature effects are taken into account by introducing a near-extremal black brane in the gravity dual and we assume that the metric of the gravity dual has the following general form
\begin{equation}
\label{eq:metric}
ds^2 = -G_{00} (U) dt^2 + G_{xx} (U) d\vec{x}^2 + G_{UU} (U) dU^2,
\end{equation}
where $\vec{x}=(x,y,z)$ denotes the usual spatial coordinates while $U$ is the radial direction. The metric \eqref{eq:metric} is assumed to have an asymptotic boundary at $U \rightarrow \infty$. The position of the horizon of the black brane, $U_h$, is given by the (first simple) root of $G_{00} (U) = 0$ starting from the boundary and we will also assume here that $G_{UU} (U_h)\to \infty$, with $G_{00} (U_h) G_{UU} (U_h)$ finite. The black brane temperature $T$ (which is a function of the position of the horizon, $T=T(U_h)$) corresponds to the temperature of the thermal bath in the gauge theory. Also, note that the thermal state of the gauge theory considered here is assumed to be invariant under spatial $SO(3)$ rotations and the $Q\bar{Q}$ pair is at rest in the local rest frame of the plasma.  

\section{Real part of the heavy quark potential}\label{realpartsection}

The calculation of the real part of the potential within the gauge/gravity duality is well-known and can now be found in textbooks \cite{kiritsisbook}. However, some of the formulas that appear in this calculation will be used in the determination of the imaginary part of the potential and thus, for the sake of completeness, we shall briefly review the necessary details here. The reader who is familiar with this subject may skip it and go directly to Section \ref{impartsection}.

For the rectangular Wilson loop, we choose the coordinate system shown in Fig.\ \ref{fig:wilsonloop} in which the string worldsheet coordinates are written in static gauge, $X^\mu = (t,x,0,0,U(x,t))$, $\tau = t$ and $\sigma = x$. Furthermore, since $\mathcal{T} \to \infty$, any slice of the worldsheet with constant $t$ 
has the same form (as shown in Fig.\ \ref{fig:loopads}(b)) - this means that we can take $U(x,t)=U(x)$. We present a sketch of a fixed $t$ slice of the string worldsheet in Fig.\ \ref{fig:loopadscorte}. With these choices and the general metric \eqref{eq:metric}, the action \eqref{eq:nambugoto} takes the form
\begin{equation}
\label{eq:nambugotostatic}
S_{NG} = \frac{\mathcal{T}}{2\pi \alpha'} \int\limits_{-L/2}^{L/2} dx \sqrt{M(U(x)) (U')^2 + V(U(x))},
\end{equation}
where $U' \equiv dU/dx$, $M(U) \equiv G_{00} G_{UU}$ and $V(U) \equiv G_{00} G_{xx}$. For the models considered in this work, we will always have $M(U) > 0$. The action 
\eqref{eq:nambugotostatic} is only implicitly dependent on $x$ and, thus, the associated Hamiltonian is a constant of motion
\begin{equation}
\label{eq:firstintegral}
\mathcal{H}_{NG} = \frac{V(U)}{\sqrt{M(U) (U')^2 + V(U)}} = \mathrm{const.} = \sqrt{V(U_*)},
\end{equation}
where $U_* \equiv U(x=0)$ and also $U'(0) = 0$ (since the string has its minimum at $x=0$). We can solve \eqref{eq:firstintegral} for $U'$ and obtain
\begin{equation}
\label{eq:dzdx}
\frac{dU}{dx} = \left[ \frac{V(U)}{M(U)} \left( \frac{V(U)}{V(U_*)}-1 \right) \right]^{1/2}.
\end{equation}
Since the endpoints of the string are located at $x=-L/2$ and $x=L/2$, we integrate \eqref{eq:dzdx} to obtain a relation between $U_*$ and $L$,
\begin{equation}
\label{eq:equationforL}
\frac{L}{2} = \int\limits^{\infty}_{U_*} dU \, \sqrt{M(U)} \left[ V(U) \left( \frac{V(U)}{V(U_*)}-1 \right) \right]^{-1/2}.
\end{equation}
We may deduce another consequence of \eqref{eq:dzdx} which will be useful later. In fact, differentiating \eqref{eq:dzdx} with respect to $x$ and then setting $x=0$ and $U=U_*$ one finds
\begin{equation}
\label{eq:upp0}
U''(0) = \frac{1}{2} \frac{V'(U_*)}{M(U_*)},
\end{equation}
where $V'(U) = dV(U)/dU$. Since $x=0$ is a minimum, $U''(0) > 0$, and one can see that $V'(U_*) > 0$.

Finally, we use \eqref{eq:dzdx} to obtain an expression for the action \label{eq:nambugotostatic} evaluated at the classical solution of the equations of motion
\begin{equation}
\label{eq:actionnotreg}
S_{str} = \frac{\mathcal{T}}{\pi \alpha'} \int\limits_{U_*}^{\infty} dU \, \sqrt{M(U)} \sqrt{\frac{V(U)}{V(U_*)}}  \left(\frac{V(U)}{V(U_*)}-1 \right)^{-1/2}.
\end{equation}
The (yet to be regularized) real part of the heavy quark potential is simply given by $\lim_{\mathcal{T}\to \infty}S_{str}/\mathcal{T}$. The equations \eqref{eq:equationforL} and \eqref{eq:actionnotreg} (minus the regularization) solve the problem. To obtain ${\rm Re}\,V_{Q\bar{Q}}$ as a function of $L$ and $T$ we either eliminate $U_*$ from both equations or, when this is not possible, parametrize both $L$ and ${\rm Re}\,V_{Q\bar{Q}}$ as functions of $U_*$. 

Note that \eqref{eq:actionnotreg} is UV divergent. This UV divergence, which is characteristic of Wilson loops, appears in the holographic approach from the fact that the string must stretch from the bulk to the boundary. Note that this is the same type of UV divergence found in $\mathcal{N}=4$ SYM at $T=0$ \cite{Maldacena:1998im}, which is to be expected since in thermal gauge theories all UV divergences must come from the vacuum contribution \cite{kapusta}. This implies that the same regularization chosen for the vacuum can be used to render the $T\neq0$ potential finite. The regularized real part of the potential at nonzero temperature can be written as
\begin{eqnarray}
\label{eq:potentialreg}
{\rm Re}\,V^{reg}_{Q\bar{Q}}(L,T)&=& \frac{1}{\pi \alpha'} \int_{U_*}^{\infty} dU \,\left[\sqrt{M(U)} \sqrt{\frac{V(U)}{V(U_*)}} \left(\frac{V(U)}{V(U_*)}-1 \right)^{-1/2}-\sqrt{M_0(U)}\right]\nonumber \\ &-& \frac{1}{\pi \alpha'}\int_0^{U_*}dU\,\sqrt{M_0(U)}.
\end{eqnarray}
where $M_0(U)=\lim_{U\to \infty}M(U)$. This temperature independent regularization scheme for the real part of the potential is well defined for any asymptotically $AdS_5$ geometry, even in the case in which the dual gauge theory displays confinement at $T=0$ (in the sense of an area law for the rectangular Wilson loop in the vacuum)\footnote{As explained in \cite{Bak:2007fk}, the regularization scheme involving the subtraction of the contribution coming from two ``straight'' strings running from $U_h$ to $U\to \infty$ is temperature dependent. Moreover, since the connected U-shaped contribution to the potential is of order $N_c^0$ and this kind of disconnected contribution involving the two straight strings is of order $N_c^2$ \cite{Bak:2007fk}, it becomes problematic to use the latter to regularize the heavy quark potential in the large $N_c$ limit where these classical gravity calculations are performed. Therefore, in this paper we opted to use the expression in Eq.\ \eqref{eq:potentialreg}, which is well defined in the large $N_c$ limit.}. 

The expectation value of the Polyakov loop $|\langle {\rm tr\,}\bold{L} \rangle|$ can be easily extracted from \eqref{eq:potentialreg} by assuming that when $L\to \infty$, $U_* \to U_h$ and ${\rm Re}\,V^{reg}_{Q\bar{Q}}\to 2F^{reg}_Q$. This gives the (regularized) heavy quark free energy
\begin{eqnarray}
\label{eq:polyloopreg}
F^{reg}_{Q}(T)&=& \frac{1}{2\pi \alpha'} \int_{U_h}^{\infty} dU \,\left[\sqrt{M(U)}-\sqrt{M_0(U)}\right]- \frac{1}{2\pi \alpha'}\int_0^{U_h}dU\,\sqrt{M_0(U)}
\end{eqnarray}
and the Polyakov loop $|\langle {\rm tr\,}\bold{L}(T) \rangle|=\exp\{-F^{reg}_{Q}(T)/T\}$. While this simple procedure gives the correct expression for $F^{reg}_{Q}(T)$ \cite{Noronha:2009ud,Noronha:2010hb} in this type of gravity duals, we note that other configurations for the string worldsheet besides the U-shaped one must be taken into account when $LT > 1$ \cite{Bak:2007fk,Grigoryan:2011cn}. In the following we will always consider the regularized expressions for the quantities discussed above and, thus, the superscript ``$reg$" will be omitted from the formulas in the rest of the text.

For further use, let us also recall the properties that the background metric must display in order for the rectangular Wilson loop to display an area law at $T=0$ \cite{Kinar:1998vq,Sonnenschein:1999if}. For the general metric in Eq.\ \eqref{eq:metric}, it was shown in \cite{Kinar:1998vq,Sonnenschein:1999if} that if there is a $U_0$ such that $V(U)$ has a minimum or $M(U_0)$ diverges (with $V(U_0) \neq 0$), then the theory linearly confines with string tension $\sigma = \frac{1}{2\pi \alpha'} \sqrt{V(U_0)}$. As one pulls the quarks apart and $L\to \infty$, the bottom of the classical string becomes flat at $U_0$ and cannot penetrate any further into the geometry. In the deconfined phase of a ($T=0$ confining) gauge theory, however, $U_0$ is hidden by the horizon and $\sigma=0$. 

\section{Thermal worldsheet fluctuations and the imaginary part of the heavy quark potential in strongly coupled plasmas}\label{impartsection}

We now generalize the procedure proposed in \cite{Noronha:2009da} to extract the imaginary part of heavy quark potential, $\mathrm{Im} \, V_{Q\bar{Q}}$, using the gauge/gravity correspondence. After deriving a formula for $\mathrm{Im} \, V_{Q\bar{Q}}$ using the saddle point approximation, we discuss its limitations and present some general conditions for the existence of such an imaginary part in this setup. We remark that other approaches have been proposed to extract the imaginary part of the potential using holography in \cite{Albacete:2008dz,Hayata:2012rw}. These different methods give results that are qualitatively equivalent in the case of $\mathcal{N}=4$ SYM theory. The method discussed in detail in this section has the advantage of being of easy implementation in comparison to the other schemes since $\mathrm{Im} \, V_{Q\bar{Q}}$ for a generic gravity dual \eqref{eq:metric} can be directly computed using the formula in Eq.\ \eqref{eq:imvqq} derived below.

\subsection{The saddle point approximation}

\begin{figure}[h!t]
\begin{center}
\includegraphics[width=0.5 \textwidth]{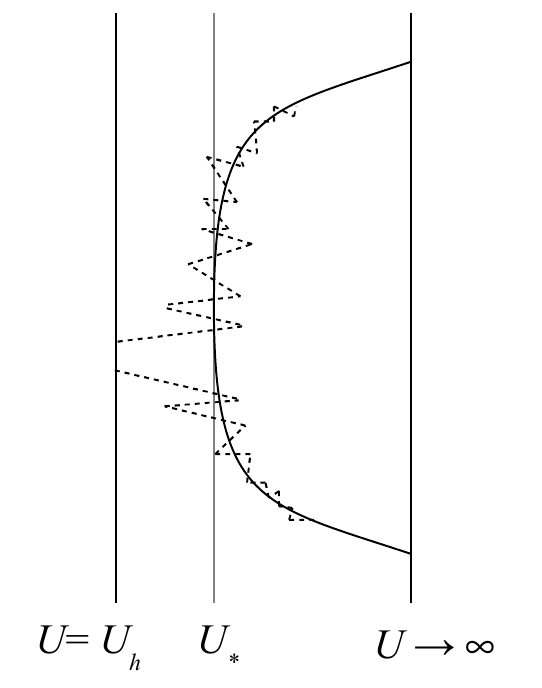}
\caption{An illustration of the effect of thermal fluctuations (dashed line) around the classical string configuration (solid line). If the bottom of the classical string solution is close enough to the horizon, thermal worldsheet fluctuations of very long wavelength may be able to reach the black brane horizon at $U_h$.}
\label{fig:thermalfluc}
\end{center}
\end{figure}

In the previous section, we saw that the classical solution to the Nambu-Goto action \eqref{eq:nambugoto} can be used to compute the real part of the heavy quark potential. To extract $\mathrm{Im} \, V_{Q\bar{Q}}(L,T)$ we have to consider the effect of thermal worldsheet fluctuations about the classical configuration $U = U_c(x)$. Such fluctuations, although taken here to be small, may turn the integrand of \eqref{eq:nambugotostatic} negative near $x=0$ and generate an imaginary part for the effective string action. The corresponding physical picture is that some part of the string, through thermal fluctuations, may reach the horizon (see Fig.\ \ref{fig:thermalfluc}).

Therefore, we shall consider the effect of worldsheet fluctuations $\delta U(x)$ ($\delta U(\pm L/2)=0$) around the classical configuration $U_c(x)$
\begin{equation}
\label{eq:thermalfluc}
U(x) = U_c(x) \rightarrow U(x) = U_c(x) + \delta U (x).
\end{equation}
The classical configuration $U_c(x)$ solves $\delta S_{NG} = 0$. For simplicity, the fluctuations $\delta U(x)$ are taken to be of arbitrarily long wavelength, i.e., $\frac{d \,\delta U(x)}{dx} \to 0$. The string partition function that takes into account the fluctuations is then
\begin{equation}
\label{eq:thermalpartition1}
Z_{str} \sim \int \mathcal{D} \delta U(x) e^{i S_{NG} (U_c(x) + \delta U (x))}.
\end{equation}
If $\delta U(x)$ is such that the integrand in $S_{NG}$ acquires an imaginary part then, by considering \eqref{eq:wilsonrec} through \eqref{eq:classicalpartition}, $\mathrm{Im} \, V_{Q\bar{Q}}(L,T) \neq 0$. Note that we are assuming that the fluctuations are not strong enough to allow for transitions to different classical extrema of $Z_{str}$.

We proceed by dividing the interval $-L/2 < x < L/2$ into $2N$ points $x_j=j \Delta x$ ($j=-N,-N+1,\ldots,N$) with $\Delta x \equiv L/(2N)$ and then take the $N\to \infty$ limit in the end of the calculation. Then, $Z_{str}$ becomes
\begin{equation}
\label{eq:thermalpartition2}
Z_{str} \sim \lim_{N\to \infty}\int d [\delta U(x_{-N})] \ldots d[ \delta U(x_{N})]  \exp{\left[ i \frac{\mathcal{T} \Delta x}{2 \pi \alpha'} \sum_j \sqrt{M(U_j) (U'_j)^2 + V(U_j)}\right]},
\end{equation}
where $U_j \equiv U(x_j)$ and $U'_j \equiv U'(x_j)$. The thermal fluctuations are most important around $x=0$ where $U=U_*$, which means that it is reasonable to expand $U_c(x_j)$ around $x=0$ and keep only terms up to second order in $x_j$. Since $U_c'(0) = 0$ we have that
\begin{equation}
\label{eq:Ucexpansion}
U_c(x_j) \approx U_* + \frac{x_j^2}{2} U_c''(0).
\end{equation}
Next, as we are considering only small fluctuations around the classical configuration, we expand $V(U(x_j))=V(U_c(x_j) + \delta U (x_j))$ in $x_j$ and $\delta U$, keeping only the terms up to second order in the monomial $x_j^m \delta U^n$
\begin{equation}
\label{eq:Vexpansion}
V(U_j) \approx V_* + \delta U V'_* + U_c''(0) V'_* \frac{x_j^2}{2} + \frac{\delta U^2}{2} V''_*,
\end{equation}
where $V_* \equiv V(U_*)$, $V'_* \equiv V'(U_*)$, and etc. The function $M(U)$ admits the same expansion as $V(U)$ but, in the action (3.1), $M(U)$
 appears only via $M(U)(U'(x))^2$. Using \eqref{eq:Ucexpansion} we see that $U'(x) \approx x_j U_c''(0)$ and, therefore, $U'(x)^2$ is already a term of second order in $x_j^m \delta U^n$. Then, we consider only the zeroth order term in the expansion of $M(U)$, i.e., $M(U) \approx M(U_*)$. Combining this with Eqs.\ 
\eqref{eq:Ucexpansion} and \eqref{eq:Vexpansion} we can approximate the exponent in \eqref{eq:thermalpartition2} as
\begin{equation}
\label{eq:approxaction}
S^{NG}_j = \frac{\mathcal{T} \Delta x}{2 \pi \alpha'} \sqrt{C_1 x_j^2 + C_2}
\end{equation}
where
\begin{equation}
\label{eq:coef1}
C_1 = \frac{U_c''(0)}{2} \left[ 2 M_* U_c''(0) + V_*' \right]
\end{equation}
and
\begin{equation}
\label{eq:coef2}
C_2 = V_* + \delta U V'_* + \frac{\delta U^2}{2} V''_*,
\end{equation}
where we defined $M_* \equiv M(U_*)$. Since $U_c''(0)>0$ and $M, V'_* > 0$, one sees that $C_1 > 0$. 

If the function in the square root of \eqref{eq:approxaction} is negative then $S^{NG}_j$ contributes to $\mathrm{Im} \, V_{Q\bar{Q}}(L,T) \neq 0$. The next step consists in determining when this happens and what the corresponding contribution to $Z_{str}$ is. In order to do that, let us isolate the $j$-th such contribution to $Z_{str}$
\begin{equation}
\label{eq:jthpartition}
I_j \equiv \int\limits_{\delta U_{j min}}^{\delta U_{j max}} d(\delta U_j) \, \exp{\left[ i \frac{\mathcal{T} \Delta x}{2 \pi \alpha'} \sqrt{C_1 x_j^2 + C_2} \right]},
\end{equation}
where $\delta U_{j min}$, $\delta U_{j max}$ are the roots of $C_1 x_j^2 + C_2$ in $\delta U$. For $\delta U_{j min} < \delta U < \delta U_{j max}$ we have $C_1 x_j^2 + C_2 < 0$, which means that \eqref{eq:jthpartition} is exactly the contribution to $\mathrm{Im} \, V_{Q\bar{Q}}(L,T) \neq 0$ we were looking for - the total contribution for all $x_j$ is $\prod_j I_j$.

The integral in \eqref{eq:jthpartition} can be evaluated using the saddle point method in the classical gravity approximation where $\alpha' \ll 1$. The exponent has a stationary point when the function
\begin{equation}
\label{eq:integrandexp}
D(\delta U_j) \equiv C_1 x_j^2 + C_2(\delta U_j)
\end{equation}
assumes an extremal value. This happens for
\begin{equation}
\label{eq:extremedeltaz}
\delta U = - \frac{V'_*}{V''_*}.
\end{equation}
Requiring that the square root has an imaginary part implies that $D(\delta U_j) < 0 \to -x_c < x_j <x_c$ where
\begin{equation}
\label{eq:criticalxj}
x_c = \sqrt{\frac{1}{C_1}\left[\frac{V'^2_*}{2V''_*} - V_* \right]}.
\end{equation}
We take $x_c = 0$ if the square root in \eqref{eq:criticalxj} is not real. Under these conditions, we can approximate $D(\delta U)$  by $D(- V'_*/V''_*)$ in \eqref{eq:jthpartition}
\begin{equation}
\label{eq:saddlepoint}
I_j \sim \exp \left[ i \frac{\mathcal{T} \Delta x}{2 \pi \alpha'} \sqrt{C_1 x_j^2 + V_* - \frac{V'^2_*}{2V''_*}} \right].
\end{equation}
Since the total contribution to the imaginary part is given by $\prod_j I_j$, returning to the continuum limit and invoking the prescription \eqref{eq:wilsongaugegravity}, we find
\begin{equation}
\label{eq:imvqq-int}
\mathrm{Im} \, V_{Q\bar{Q}} = -\frac{1}{2\pi \alpha'} \int\limits_{|x|<x_c} dx \sqrt{-x^2 C_1 - V_* + \frac{V'^2_*}{2V''_*}}\,.
\end{equation}
Evaluating the integral in \eqref{eq:imvqq-int} and using \eqref{eq:upp0} and \eqref{eq:coef1} we finally find a closed expression for $\mathrm{Im} \, V_{Q\bar{Q}}$
\begin{equation}
\label{eq:imvqq}
\mathrm{Im} \, V_{Q\bar{Q}} = -\frac{1}{2 \sqrt{2} \alpha'} \sqrt{M_*} \left[\frac{V'_*}{2V''_*}-\frac{V_*}{V'_*} \right].
\end{equation}

Eq.\ \eqref{eq:imvqq} reduces to the result derived in \cite{Noronha:2009da} where it was assumed that the background metric was such that $M(U)=1$. The only difference between the general formula in \eqref{eq:imvqq} and the previous one found in \cite{Noronha:2009da} is the presence of the factor $\sqrt{M_*}$ ($M$ gives an idea of how much warped the space-time is in the bulk). Also, note that ${\rm Im}\,V_{Q\bar{Q}}$ is UV finite. Moreover, the fluctuations also change the real part of the potential. This is discussed in Appendix \ref{covariantmethod}. 

An important condition that must be satisfied in order for the saddle point calculation shown here to be applicable is $V''_* \neq 0$. If $V''_* = 0$ then \eqref{eq:integrandexp} does not have extrema and higher orders terms in $\delta U$ must be kept in the expansion \eqref{eq:Vexpansion} for $V$, which signals the breakdown of the saddle point approximation. 

Finally, an alternative derivation of the imaginary part of $V_{Q\bar{Q}}$ using a covariant background expansion of the Nambu-Goto action is given in Appendix \ref{covariantmethod}.

\subsection{The relationship between $\mathrm{Im} \, V_{Q\bar{Q}}$, confinement, and the black brane}
\label{sec:conf}

A first glance into the derivation of Eq.\ \eqref{eq:imvqq} may give the misleading idea that the presence of a black brane is not necessary in order to have $\mathrm{Im} \, V_{Q\bar{Q}}\neq 0$. However, the absence of a black brane implies that $\mathrm{Im} \, V_{Q\bar{Q}} = 0$, as we shall explain below. Moreover, when the metric satisfies the conditions for the presence of an area law for the rectangular Wilson loop mentioned in Section \ref{realpartsection} one can show that $\mathrm{Im} \, V_{Q\bar{Q}} (L\to \infty) = 0$. These results are exact within the semiclassical approximation used for the string partition function in Eq.\ \eqref{eq:thermalpartition1}.

\subsubsection*{The existence of a black brane is necessary for $\mathrm{Im} \, V_{Q\bar{Q}} \neq 0$}

As mentioned in Section \ref{generalformulasection}, for the general metric in Eq.\ \eqref{eq:metric} $G_{00}(U_h) = 0$ and $G_{UU} (U_h) \to \infty$, with $G_{00} (U_h) G_{UU} (U_h)$ being finite. Therefore, $M(U)$ is finite and positive for every $U$ and $V(U) > 0$ if $U>U_h$, but $V(U_h) = 0$. The important point here is that for $U<U_h$ it is possible that $V(U)<0$.

In \eqref{eq:thermalpartition1}, requiring that the square root possesses an imaginary part means that $K(U) \equiv M(U) (U')^2 + V(U) < 0$ for some $U$. Since $M(U),V(U)>0$ for $U>U_h$, for any worldsheet 
fluctuation $\delta U$ such that $U(x)=U_c (x) + \delta U(x) > U_h$ one has $K > 0$ for every $x \in [-L/2,L/2]$. However, if the fluctuation is such that $U < U_h$, $V(U)$ may be negative and $K(U(x)) < 0$ 
for some interval in $x$ even though $M(U)>0$. In other words, if the worldsheet fluctuations are such that a portion of the string reaches the horizon and probes the black brane, then an imaginary part for the heavy quark potential may be generated. This is illustrated in Fig.\ \ref{fig:thermalfluc}. Therefore, in this approach an imaginary part for $V_{Q\bar{Q}}$ appears when we consider worldsheet fluctuations in which $\delta U < 0$.

On the other hand, if a black brane horizon is not present and the metric \eqref{eq:metric} is regular everywhere we have that $G_{00},G_{UU}$ is positive for every $U>0$. Then, we have $M(U),V(U) > 0$ and, thus, $K(U) > 0$ for every $U>0$. This implies that $\mathrm{Im} \, V_{Q\bar{Q}} = 0$, exactly. Therefore, in our approach the heavy quark potential can develop an imaginary part due to the thermal worldsheet fluctuations induced by the presence of a black brane. 

\subsubsection*{If the rectangular Wilson loop displays an area law then $\mathrm{Im} \, V_{Q\bar{Q}} (L\to \infty) = 0$}

Suppose that $M(U)$ diverges at the confinement scale $U_0$ (with $V(U_0) \neq 0$). In this case, for large $L$ we have $U_* \sim U_0$. Moreover, in this case when $L\to \infty$ the string worldsheet lays nearly flat at $U_0$. We may 
write $U_c (x) \sim U_0 -\epsilon$, where $\epsilon \ll U_0$. Finally, since $M(U_0)$ is large we may neglect the second term in the expression of $K(U)$. Therefore, for long wavelength fluctuations $\delta U' = 0$ the Nambu-Goto action in \eqref{eq:thermalpartition1} takes the form
\begin{equation}
\label{eq:actionMdiv}
S_{NG} \approx \frac{\mathcal{T}}{2\pi \alpha'} \int \limits^{L/2}_{-L/2} dx \, U_0 \sqrt{M(U_0 - \epsilon + \delta U)}.
\end{equation}
Note that now we cannot consider fluctuations such that $\delta U > \epsilon$ since then we would be taking $M$ past its divergence. Therefore, only fluctuations with $\delta U < \epsilon$ are allowed in this case. However, note that this implies that $M>0$ and, thus, the square root that appears in the evaluation of the potential is always real. Therefore, in this situation $\mathrm{Im} \, V_{Q\bar{Q}} = 0$.

Alternatively, suppose now that $M$ does not diverge at $U_0$ but rather that $V(U)$ has a minimum at $U_0$. For small fluctuations about $U_c(x) = U_0$ where $U_c'(0) = U_c''(0) = \ldots = 0$ (since the string lays nearly flat at $U_0$) one finds
\begin{equation}
\label{eq:Vexpansion2}
V(U_0+\delta U) \approx V(U_0) + \frac{1}{2} V''(U_0) \delta U^2,
\end{equation}
where $V''(U_0)>0$. Thus, $V(U)>0$ in the neighborhood of $U=U_0$, $x=0$ and $K(U)>0$. Therefore, $S_{NG}$ is real and $\mathrm{Im} \, V_{Q\bar{Q}} = 0$. We then conclude that $\mathrm{Im} \, V_{Q\bar{Q}} (L\to \infty) = 0$ if the background metric is such that the rectangular Wilson loop displays an area law. 

\begin{figure}[h!t]
\begin{center}
\includegraphics[width=0.8 \textwidth]{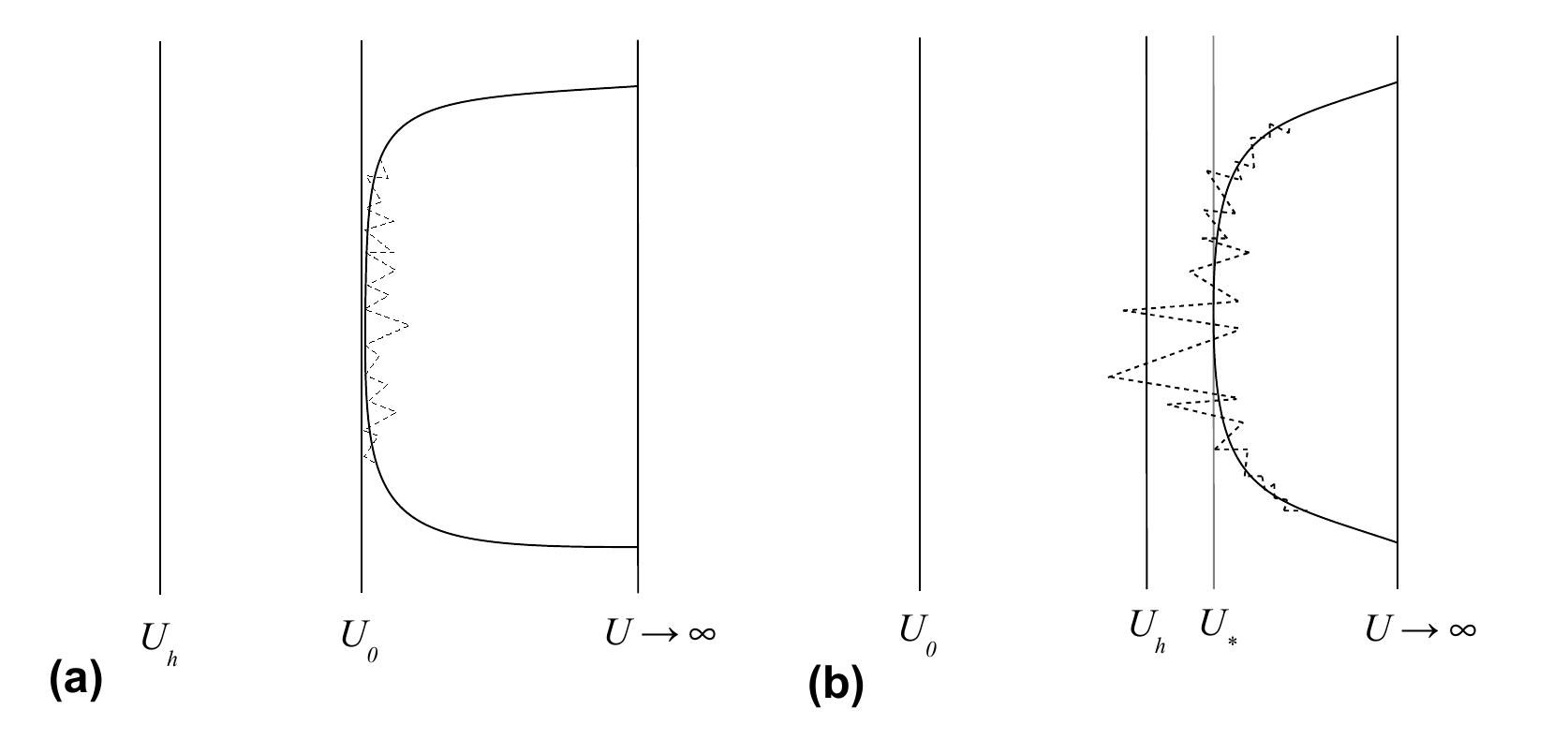}
\caption{An illustration of the relationship between thermal worldsheet fluctuations and confinement. Fig.\ \ref{fig:thermalflucconf}(a) shows that when $U_0>U_h$ the classical string worldsheet, even with 
the inclusion of thermal fluctuations, cannot go beyond $U_0$. On the other hand, Fig.\ \ref{fig:thermalflucconf}(b) shows that when $U_0 < U_h$ the horizon hides the ``barrier'' at $U_0$ and the thermal fluctuations can reach the horizon.}
\label{fig:thermalflucconf}
\end{center}
\end{figure}

We may summarize these results as follows. Suppose that $U_0$ is the value of the $U$ coordinate at which the metric satisfies the conditions for confinement and that $U_h$ is the position of the black brane horizon. If $U_0>U_h$ then the classical string cannot go past $U_0$. As discussed above, we cannot consider fluctuations beyond $U_0$. Effectively, $U_0$ acts as a ``barrier'' for the classical string. However, if $U_h>U_0$, the horizon hides this barrier and we may have fluctuations that reach $U_h$. Both cases are sketched in Fig. \ref{fig:thermalflucconf}. 

\subsection{Using $\mathrm{Im} \, V_{Q\bar{Q}}$ to estimate the thermal width of heavy quarkonia at strong coupling}

In the next section we will compute $\mathrm{Im} \, V_{Q\bar{Q}}$ in two different conformal plasmas using the prescription derived above. To estimate the thermal width $\Gamma_{Q\bar{Q}}$ of the heavy $Q \bar{Q}$ pair we will use a first-order non-relativistic expansion
\begin{equation}
\label{eq:thermalwidth}
\Gamma_{Q\bar{Q}} = - \langle \psi | \mathrm{Im} \, V_{Q\bar{Q}}(L,T) | \psi \rangle,
\end{equation}
where 
\begin{equation}
\label{eq:wavefunction}
\langle \vec{r} |\psi \rangle = \frac{1}{\sqrt{\pi} a_0^{3/2}} e^{-r/a_0}
\end{equation}
is the ground-state wave function of a particle in a Coulomb-like potential of the form $V(L) = - K/L$ and $a_0 = 2/(m_Q K)$ is the Bohr radius ($m_Q$ is the mass of the heavy quark $Q$ such that $m_Q/T \gg 1$). Even though the real part of the potential at finite temperature for the cases studied here is not given by just the $\sim 1/L$ term, this provides the leading contribution for the potential between deeply bound $Q\bar{Q}$ states in a conformal plasma, which justifies the use of Coulomb-like wave functions to determine the width. Moreover, in potential models of the bottomonium spectrum, the $\Upsilon (1S)$ state is mostly bound due to the Coulomb part of the Cornell potential. The thermal width is then given by
\begin{equation}
\label{eq:thermalwidth2}
\Gamma_{Q\bar{Q}} =-\frac{4}{a_0^3} \int_0^{\infty} dL \, L^2 e^{-2L/a_0}\, \mathrm{Im} \, V_{Q\bar{Q}}(L,T)\,.
\end{equation}
Actually, as it will be discussed shortly, we should take \eqref{eq:thermalwidth2} as representing a lower bound for the heavy quarkonia thermal width computed within the thermal worldsheet fluctuation method presented here.

\section{Calculation of $\mathrm{Im} \, V_{Q\bar{Q}}$ in some gravity duals}
\label{models}

\subsection*{An overview of the models}

Using the general framework described in the previous section, we shall now study the imaginary part of the heavy quark potential and the corresponding heavy quarkonia thermal width in two different strongly coupled plasmas dual to theories of classical gravity. In particular, we will consider the following models:

\begin{enumerate}
	\item Strongly coupled, thermal $\mathcal{N}=4$ SYM at large $N_c$. This case was already studied in \cite{Noronha:2009da} but here we shall perform a more complete study of the imaginary part of the potential and revisit the estimate for the thermal width of heavy quarkonia at strong coupling done in \cite{Noronha:2009da}.
	
	
	\item Gauss-Bonnet gravity \cite{Zwiebach:1985uq,Buchel:2008vz,Buchel:2009sk}. This model includes $\mathcal{R}_{\mu\nu\lambda\rho}^2$ terms in the gravity dual action corresponding to higher order derivative corrections to the supergravity action. 
		
\end{enumerate}

In Appendix \ref{sec:dpstacks} we discuss some other results involving Wilson loops and compute $\mathrm{Im} \, V_{Q\bar{Q}}$ for simple models of non-conformal strongly plasmas.

\subsection{$\mathcal{N}=4$ SYM}

The metric for a near-extremal black-brane in $AdS_5 \times S_5$ is given by
\begin{equation}
\label{eq:metricads}
ds^2 = -\frac{U^2}{R^2} f(U) dt^2 + \frac{U^2}{R^2} d\vec{x}^2 + \frac{R^2}{U^2} \frac{1}{f(U)} dU^2 + R^2 d\Omega_5^2,
\end{equation}
where $R$ is the common radius of $S_5$ and $AdS_5$, $f(U) \equiv 1 - U_h^4/U^4$, $d\Omega^2_5$ corresponds to the $S_5$ part of metric and, as before, $U_h$ is the position of the black brane horizon. The boundary gauge theory is $\mathcal{N} = 4$ SYM with $N_c\to \infty$. The 't Hooft coupling in this strongly coupled gauge theory is given by $\lambda = R^4/\alpha'^2 \gg 1$. The temperature of the black brane (and of the dual gauge theory) is given by 
\begin{equation}
\label{eq:tempAdS}
T = \frac{U_h}{\pi R^2}.
\end{equation}
In the following we always choose a fixed configuration for the string coordinates in $S_5$ and, thus, all the calculations are effectively done only using the $AdS_5$ piece. For this metric $M(U) = 1$ and $V(U) = (U^4-U_h^4)/R^4$.

\subsection*{Heavy quark potential in the vacuum}

The expressions for $L$ \eqref{eq:equationforL} and $S^{nreg}$ \eqref{eq:actionnotreg} turn, in this case, into
\begin{equation}
\label{eq:LAdST0}
\frac{L}{2} = \frac{R^2}{U_*} \int \limits_{1}^{\infty} dy \, \frac{1}{y^2\sqrt{y^4-1}} \quad \text{and}
\end{equation}
\begin{equation}
\label{eq:SnregAdST0}
S^{nreg} = \frac{\mathcal{T}}{\pi \alpha'} U_* \int \limits_{1}^{\infty} dy \, \frac{y^2}{\sqrt{y^4-1}}.
\end{equation}
where we made the change of variables $U \to y = U/U_*$. Note that the integral \eqref{eq:SnregAdST0} diverges linearly when $y \to \infty$ and this is the UV divergence we already expected. The regularized potential is
\begin{equation}
\label{eq:VQQregAdST0}
V_{Q\bar{Q}} = \frac{U_*}{\pi \alpha'} \left[\int \limits_{1}^{\infty} dy \, \left( \frac{y^2}{\sqrt{y^4-1}}-1 \right) -1 \right].
\end{equation}

The integrals in \eqref{eq:LAdST0} and \eqref{eq:VQQregAdST0} can be done in terms of the beta function, as described in Appendix \ref{sec:integrals}. After integration, one obtains
\begin{equation}
\label{eq:LU0AdST0}
L = \frac{R^2}{U_*} \frac{2 \sqrt{2\pi} \pi}{\Gamma(1/4)^2} 
\end{equation}
and
\begin{equation}
\label{eq:VQQU0AdST0}
V_{Q\bar{Q}} = - U_* \frac{\sqrt{2\pi}}{\alpha' \Gamma(1/4)^2}.
\end{equation}
In this particular case, it is possible to eliminate the parameter $U_*$ from \eqref{eq:LU0AdST0} and \eqref{eq:VQQU0AdST0} to obtain the potential as an explicit function of $L$ \cite{Maldacena:1998im}
\begin{equation}
\label{eq:VQQAdST0-final}
V_{Q\bar{Q}} = -\frac{4 \pi^2}{\Gamma(1/4)^4} \frac{R^2}{\alpha'} \frac{1}{L}\,.
\end{equation}

From \eqref{eq:VQQAdST0-final} we obtain an estimate for the Bohr radius that will be used throughout this work, $a_0 = \Gamma(1/4)^4/(m_Q 2 \pi^2 \sqrt{\lambda})$. For the case of a bottom quark $m_b \sim 4.7$ GeV and, using $\lambda = 9$ \cite{Noronha:2009vz}, one finds $a_0 \sim 0.6 \, \mathrm{GeV^{-1}}$.

\subsection*{Thermal $\mathcal{N}=4$ SYM}

We start by computing the heavy quark free energy from Eq.\ \eqref{eq:polyloopreg}. For its regularization we use half of the regularization term used for the potential at $T=0$, which gives 
\begin{equation}
\label{eq:FQAdSTneq0}
F_Q = -\frac{U_h}{2 \pi \alpha'}\,.
\end{equation}
Using \eqref{eq:tempAdS} we can write \eqref{eq:FQAdSTneq0} as
\begin{equation}
\label{eq:FQAdSTneq0}
\frac{F_Q}{T} = -\frac{\sqrt{\lambda}}{2}.
\end{equation}
This result \cite{Brandhuber:1998bs,Rey:1998bq} is consistent with the fact that the only scale available in the calculation of the Polyakov loop in a thermal $\mathcal{N}=4$ SYM theory is the temperature $T$.

For the rectangular Wilson loop at finite $T$ we use the same regularization employed for the $T=0$ case. The resulting expressions for $L$ and ${\rm Re}\,V_{Q\bar{Q}}$ may be written as \cite{Brandhuber:1998bs,Rey:1998bq}
\begin{equation}
\label{eq:LTAdSTneq0}
LT(y_h) = \frac{2}{\pi} y_h \sqrt{1-y_h^4} \int\limits^{\infty}_1 \frac{dy}{\sqrt{(y^4-y_h^4)(y^4-1)}}
\end{equation}
\begin{equation}
\label{eq:FQQAdSTneq0}
\frac{{\rm Re}\,V_{Q\bar{Q}}(y_h)}{T} = \frac{R^2}{\alpha'} \frac{1}{y_h} \left[ \int\limits_1^{\infty} dy \,\left( \sqrt{\frac{y^4-y_h^4}{y^4-1}} - 1 \right) - 1 \right]
\end{equation}
where $y_h \equiv U_h/U_*$ and $0 < y_h < 1$. 

These integrals can be calculated in terms of hypergeometric functions \cite{Albacete:2008dz} as shown in Appendix \ref{sec:integrals}. One finds that
\begin{equation}
\label{eq:LTAdSTneq0-res}
LT(y_h) = \frac{2 \sqrt{2\pi}}{\Gamma(1/4)^2} y_h \sqrt{1-y_h^4} \, {}_2 F_1\left[\frac{1}{2},\frac{3}{4};\frac{5}{4};y_h^4 \right] \quad \text{and}
\end{equation}
\begin{equation}
\label{eq:FQQAdSTneq0-res}
\frac{{\rm Re}\,V_{Q\bar{Q}}}{T} = - \frac{R^2} {\alpha'} \frac{\sqrt{2\pi^3}}{\Gamma(1/4)^2} \frac{1}{y_h} \, {}_2 F_1 \left[-\frac{1}{2},-\frac{1}{4} ; \frac{1}{4};y_h^{4} \right].
\end{equation}
These equations cannot be solved exactly and must be analyzed as a function of $y_h$. However, when $LT \ll 1$ it is possible to expand both expressions in powers of $(LT)^4$, obtaining, to first order (Appendix \ref{sec:integrals})
\begin{equation}
\label{eq:FQQAdSTneq0approx}
\frac{{\rm Re}\,V_{Q\bar{Q}}}{T} = -\frac{4 \pi^2 \sqrt{\lambda}}{\Gamma(1/4)^4 LT}\left[1+c(LT)^4\right],
\end{equation}
where 
\begin{equation}
\label{eq:constantapprox}
c= \frac{3}{5 \cdot 2^7 \pi^2} \Gamma(1/4)^8.
\end{equation}
The fact that the potential only depends on the combination $LT$ is expected since $\mathcal{N}=4$ SYM is a conformal plasma. 

\begin{figure}[htp!]
\begin{center}
\includegraphics[width=0.7 \textwidth]{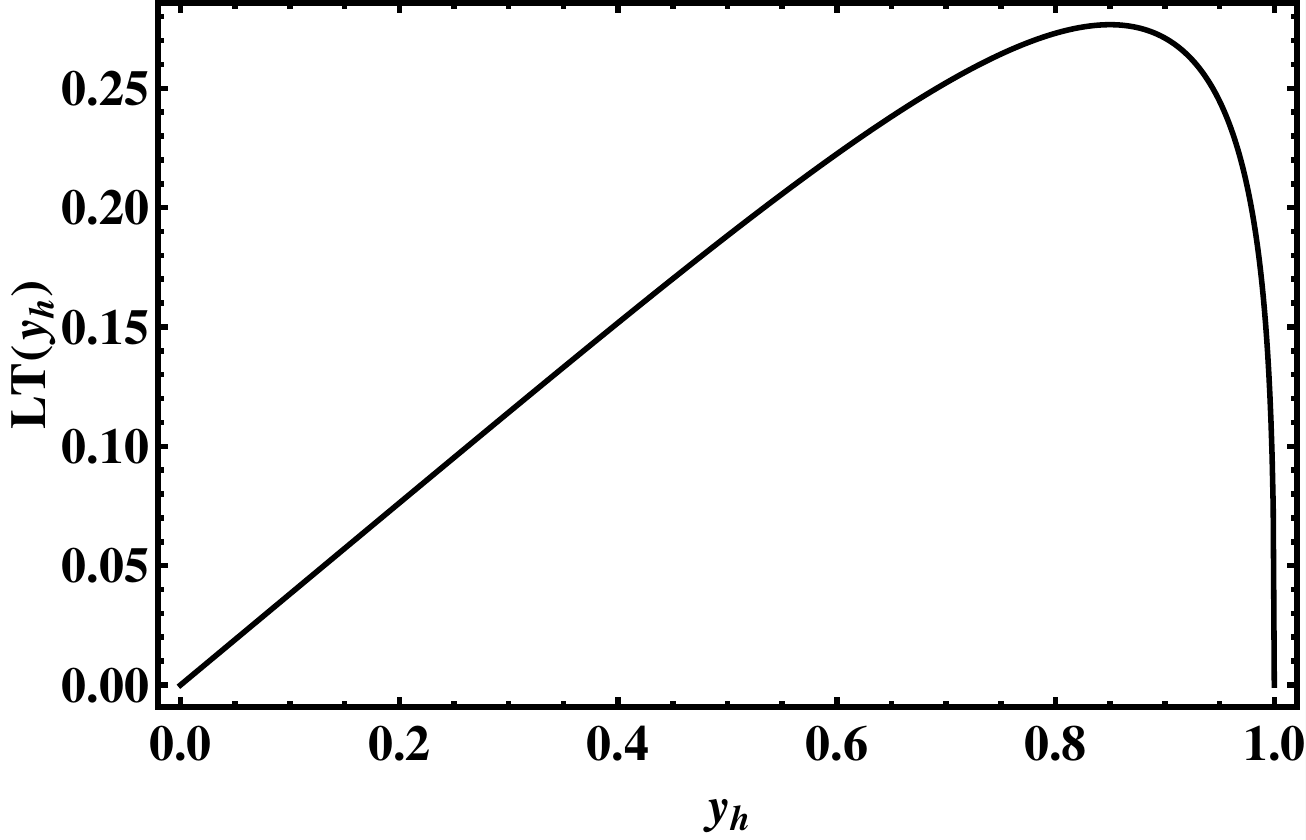}
\caption{$LT$ as a function of $y_h$ for the case of $\mathcal{N}$ = 4 SYM at strong coupling. For $y_h > y_{h,max} \sim 0.85$ the solution of the classical Nambu-Goto action is not the dominant configuration and other connected configurations must also be taken into account \cite{Bak:2007fk}.}
\label{fig:LTfuncyh}
\end{center}
\end{figure}

Let us examine \eqref{eq:LTAdSTneq0-res}. In Fig.\ \ref{fig:LTfuncyh} we plot $LT$ as a function of $y_h$. One sees that there is a maximum value of $y_h$, $y_{h,max} = 0.85$, and that $LT$ is a decreasing function of $y_h$ for $y_h > y_{h,max}$. Physically, this means that for $y_h > y_{h,max}$, one has to take into account highly curved configurations for the string worldsheet which are not solutions of the Nambu-Goto action but are important for $y_h > y_{h,max}$ \cite{Bak:2007fk}. In fact, a calculation of the curvature scalar associated with the worldsheet metric in Appendix D shows that it diverges for $y_h\to 1$. Therefore, we can only trust this U-shaped classical solution up to $y_{h,max}$. For further reference, the corresponding value of $LT$ is $LT_{max} = LT(y_{h,max}) \sim 0.28$. From Fig.\ \ref{fig:LTfuncyh} we also see that for $y_h \sim 0$, $LT \approx b y_h$, where $b =2 \sqrt{2 \pi}/\Gamma(1/4)^2 \sim 0.38$.

\begin{figure}[htp!]
\begin{center}
\includegraphics[width=0.7 \textwidth]{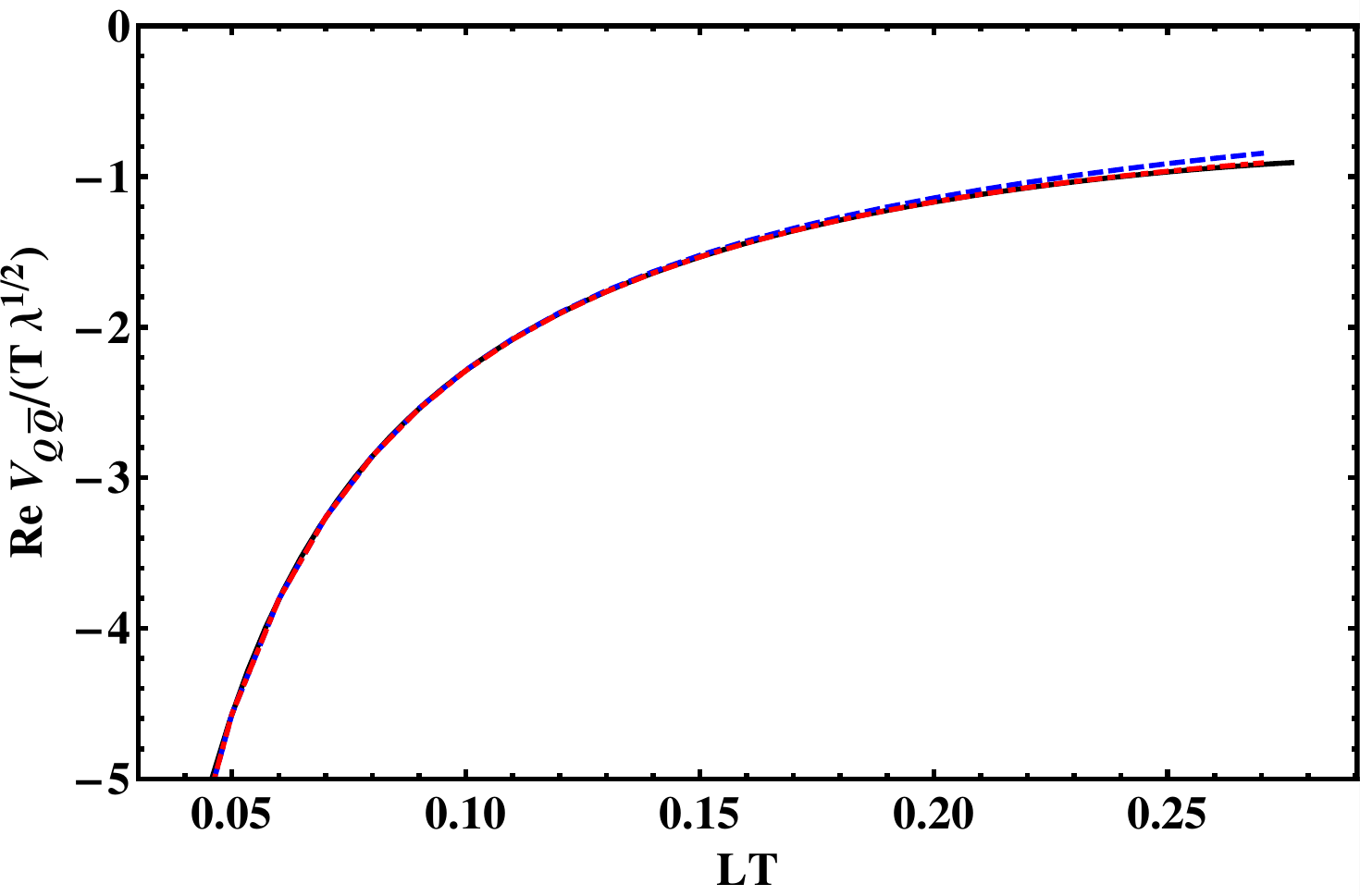}
\caption{The potential ${\rm Re}\,V_{Q\bar{Q}}/T$ for strongly coupled $\mathcal{N}$ = 4 SYM (normalized by the 't Hooft coupling $\sqrt{\lambda}$) as a function of $LT$, considering the exact solution given by \eqref{eq:LTAdSTneq0-res} and \eqref{eq:FQQAdSTneq0-res} (solid black curve), the approximation \eqref{eq:FQQAdSTneq0approx} valid for $LT \ll 1$ (dotted-dashed red curve), and the vacuum limit given by \eqref{eq:VQQAdST0-final} (dashed blue curve).}
\label{fig:grafFQQdT}
\end{center}
\end{figure}

We show in Fig.\ \ref{fig:grafFQQdT} the real part of the potential ${\rm Re}\,V_{Q\bar{Q}}/T$ computed in an analogous fashion (using only the allowed interval $0 < y_h< y_{h,max}$), along with the vacuum result \eqref{eq:VQQAdST0-final} and the $LT\ll 1$ approximation \eqref{eq:FQQAdSTneq0approx}. One can see that the vacuum contribution is very close to the thermal one. Also, the $LT \ll 1$ approximation is excellent for all values of $LT$ in the allowed interval.

\subsection*{Estimating the Debye mass}
\label{sec:adsdebyemass}

We now shall describe a way to estimate the Debye screening mass $m_D$ directly from the real part of the heavy quark potential. This approach is simple and driven primarily by phenomenological reasons. Yet, it provides results qualitatively similar to more refined estimates involving, for example, the lightest CT-odd supergravity mode \cite{Bak:2007fk}.

One may define the Debye mass $m_D(T)$ as the screening mass in the $Q\bar{Q}$ potential of the Karsch-Merh-Satz (KMS) model \cite{Karsch:1987pv}
\begin{equation}
\label{eq:kmsmodelF}
\frac{{\rm Re}\,V_{Q\bar{Q}}(L,T)}{\sqrt{\lambda}} = -\tilde{C}_1 \frac{e^{-m_D(T) L}}{L} + \frac{\sigma}{m_D}\left(1-e^{-m_D(T) L} \right)+\tilde{C}_2,
\end{equation}
where $\tilde{C}_1$ is a Coulomb coupling constant, $\tilde{C}_2$ is a constant that appears due to the regularization procedure, and $\sigma$ is the string tension (normalized by $\sqrt{\lambda}$). The model \eqref{eq:kmsmodelF} describes, for $m_D \sim 0$, a Cornell-like potential and, for $\sigma \to 0$ but $m_D \neq 0$, a Debye screened Coulomb potential. For nonzero $m_D$ and $\sigma$ the result interpolates between both limits. For a conformal field theory, such as $\mathcal{N} = 4$ SYM, we can take $\sigma = 0$. Also, we know that in such theories ${\rm Re}\,V_{Q\bar{Q}}/T$ can only depend on $LT$. With this in mind, we write \eqref{eq:kmsmodelF} in the form
\begin{equation}
\label{eq:kmsmodelF-conf}
\frac{{\rm Re}\,V_{Q\bar{Q}}}{\sqrt{\lambda}T} = -\tilde{C}_1 \,\frac{e^{-\frac{m_D}{T} (LT)}}{(LT)^\delta}+\tilde{C}_2,
\end{equation}
where $m_D/T$ must be a temperature independent constant in a conformal plasma and $\delta$ is an adjustable parameter. A similar function has been used to fit lattice data for the potential (see the review in \cite{Philipsen:2010gj}).

In the following we will use \eqref{eq:kmsmodelF-conf} to obtain an estimate for $m_D$ through a fit to the numerical results for ${\rm Re}\,V_{Q\bar{Q}}/T$ as a function $LT$. However, we must stress that this is only a very rough estimate. First, equation \eqref{eq:kmsmodelF} is only a phenomenological model for the effect of Debye screening in non-Abelian gauge theories. Second, and most importantly, the solution \eqref{eq:LTAdSTneq0-res} and \eqref{eq:FQQAdSTneq0-res} imply that ${\rm Re}\,V_{Q\bar{Q}}/T$ computed using the classical string does not show exponential screening. This can be easily seen using a property of the derivative of the hypergeometric function (as discussed in Appendix B). Nevertheless, this is a very simple way to estimate $m_D$ and moreover \eqref{eq:kmsmodelF-conf} provides a reasonable description of ${\rm Re}\,V_{Q\bar{Q}}/T$.

\begin{figure}[htp!]
\begin{center}
\includegraphics[width=0.7 \textwidth]{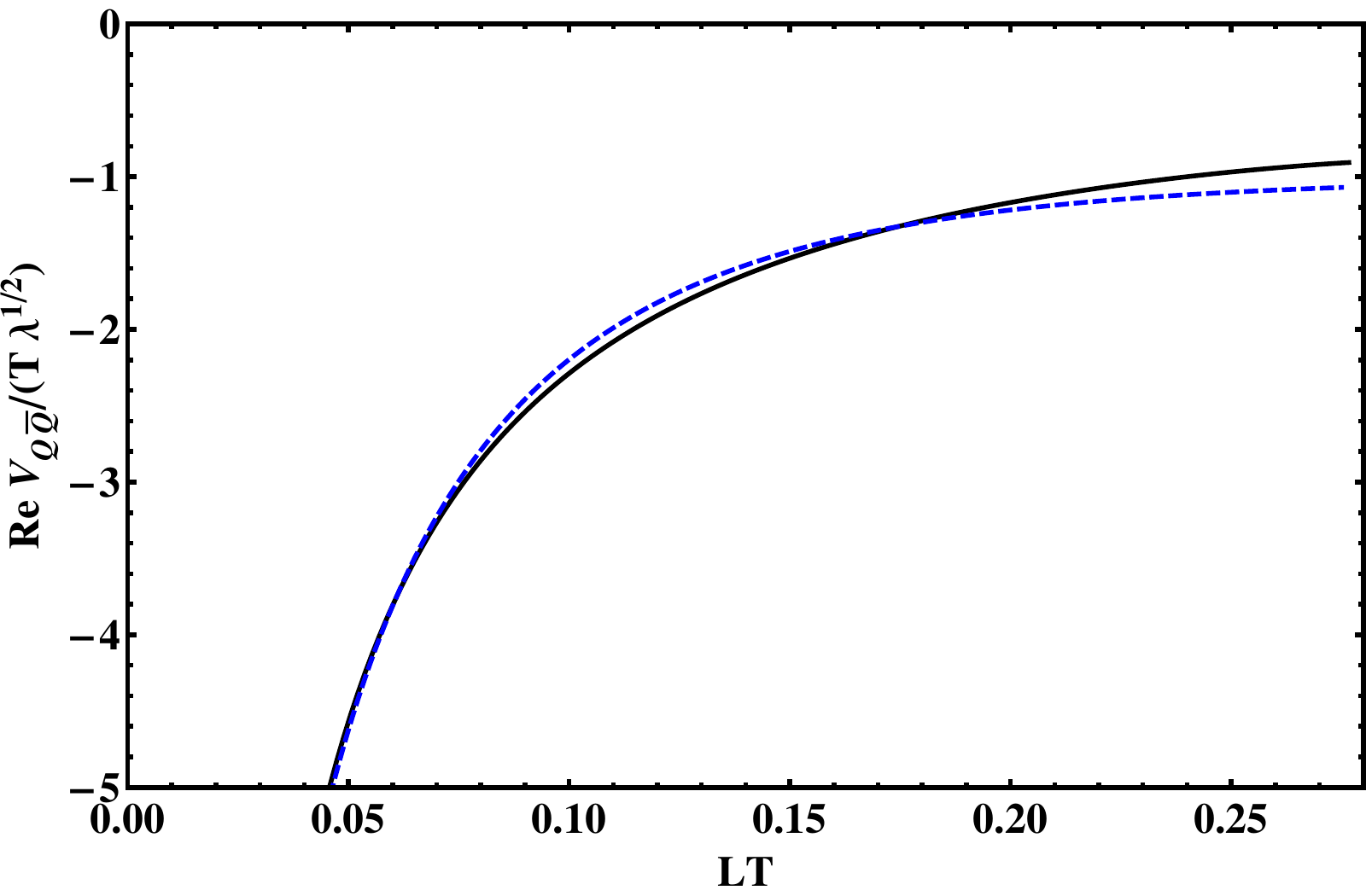}
\caption{A comparison of the exact result for ${\rm Re}\,V_{Q\bar{Q}}/T$ (solid black curve) and the fitted function \eqref{eq:kmsmodelF-conf} (dashed blue curve) for strongly coupled $\mathcal{N}$ = 4 SYM.}
\label{fig:mDN4SYM}
\end{center}
\end{figure}

The numerical procedure is to fit \eqref{eq:kmsmodelF-conf} to the exact result given by \eqref{eq:LTAdSTneq0-res} and \eqref{eq:FQQAdSTneq0-res} using $\tilde{C}_1$, $\delta$, and $m_D$ as fitting parameters ($\tilde{C}_2=-1$ is fixed by our regularization procedure). We obtain
\begin{equation}
\label{eq:debyemassAdSCFT}
m_D/T = 11.92\,\qquad \tilde{C}_1=0.72 \,, \qquad \delta=0.74\,.
\end{equation}
The exact result and the fitted function are shown in Fig.\ \ref{fig:mDN4SYM}. As a comparison, the calculation of the screening mass using the lightest CT-odd mode of type IIB supergravity gives $m_D/T = 10.694$ \cite{Bak:2007fk}.

\subsection*{Imaginary part of the heavy quark potential in $\mathcal{N}=4$ SYM}

From the general formula in Eq.\ \eqref{eq:imvqq} we obtain
\begin{equation}
\label{eq:ImVQQAdScase}
\frac{\mathrm{Im} \, V_{Q\bar{Q}}}{T} = -\frac{\pi \sqrt{\lambda}}{24 \sqrt{2}}\frac{3 y_h^4 - 1}{y_h}.
\end{equation}
The condition $\mathrm{Im} \, V_{Q\bar{Q}} <0$ implies $y_h > y_{h,min} = 3^{-1/4} \approx 0.760$. This translates into $LT > LT_{min} = 0.266$. For $LT < LT_{min}$, $\mathrm{Im} \, V_{Q\bar{Q}} =0$. As before, we can trust this solution only if $y_h < y_{h,max}$. For $y_h > y_{h,max}$ we should consider other connected contributions and the formalism developed above to determine $\mathrm{Im} \, V_{Q\bar{Q}}$ is not valid. It should also be noted that $\mathrm{Im} \, V_{Q\bar{Q}}/T$ depends only on $LT$ (via $y_h$), as expected to occur in a conformal plasma.

\begin{figure}[htp!]
\begin{center}
\includegraphics[width=0.7 \textwidth]{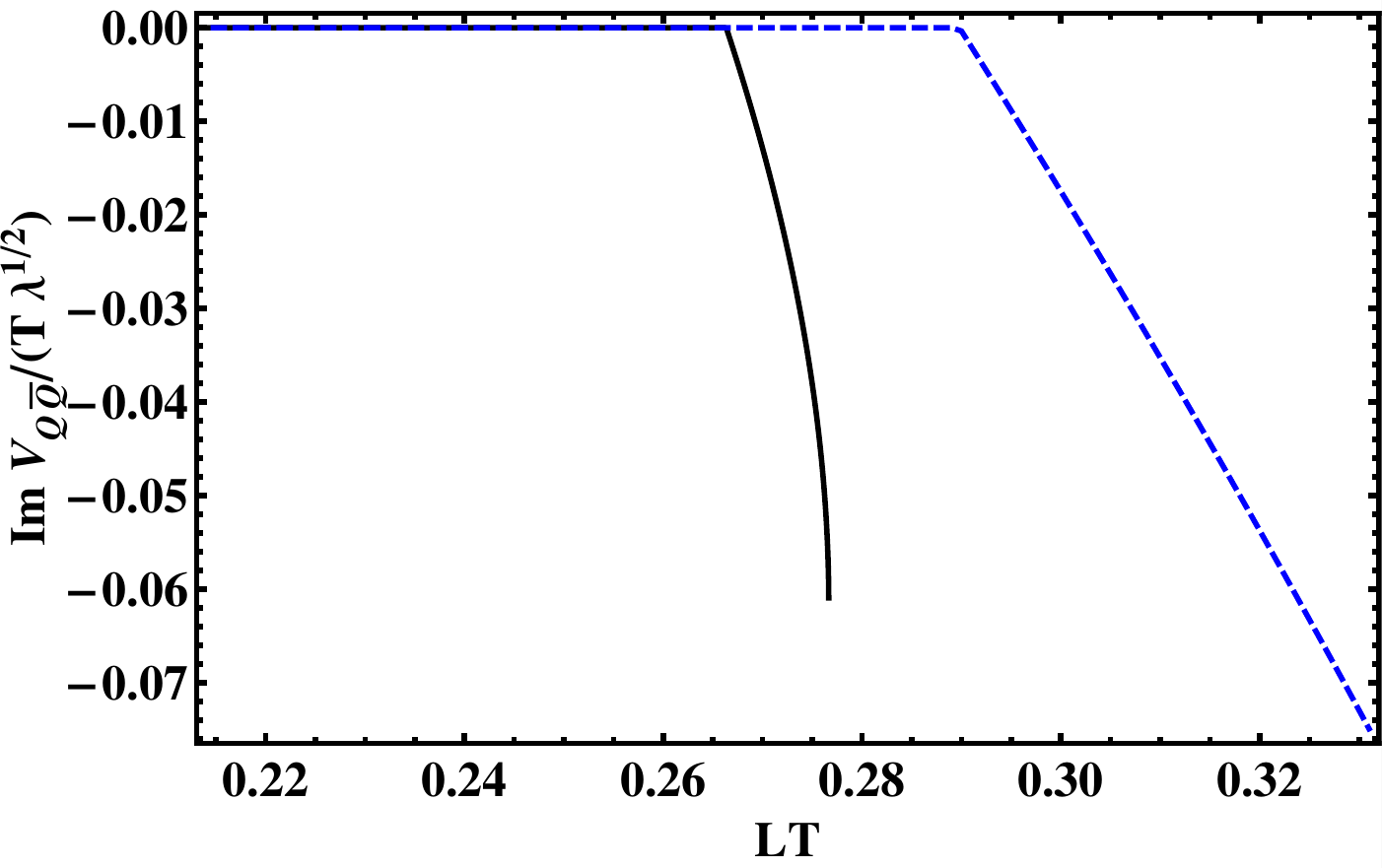}
\caption{The imaginary part of ${\rm Im}\,V_{Q\bar{Q}}/T$ as a function of $LT$. The solid black curve is the result using equation \eqref{eq:LTAdSTneq0} to eliminate $y_h$, considering only $LT < LT_{max}$. The dashed blue curve is obtained using the approximation $b y_h \sim LT$, which ignores the fact that one should not trust \eqref{eq:ImVQQAdScase} when $LT > LT_{max}$. Using this approximation, the root of \eqref{eq:ImVQQAdScase} is shifted to the right.}
\label{fig:ImVQQLT}
\end{center}
\end{figure}

One can now use \eqref{eq:LTAdSTneq0} and \eqref{eq:ImVQQAdScase} to determine the behavior of $\mathrm{Im} \, V_{Q\bar{Q}}/T$ as a function of $LT$. This is shown in Fig.\ \ref{fig:ImVQQLT} considering only $LT<LT_{max}$. We also show the result obtained using the approximation $LT \approx b y_h$, which ignores the fact that we should trust \eqref{eq:ImVQQAdScase} only for $y_h < y_{h,max}$ (in this case the root of \eqref{eq:ImVQQAdScase} is shifted to the right).

From Fig.\ \ref{fig:ImVQQLT} we conclude that we are only able to reliably calculate $\mathrm{Im} \, V_{Q\bar{Q}}$ using (5.20) in a small range of $LT$. The approximation $b y_h \sim LT$ is poor for two reasons. First, it is being used in a region of $y_h$ near $y_{h,max}$. Second, the extrapolation performed in the region $y_h > y_{h,max}$ is done beyond the trusted region for $y_h$. Nevertheless, the linear behavior of $\mathrm{Im} \, V_{Q\bar{Q}}$ seen in Fig.\ \ref{fig:ImVQQLT} agrees, qualitatively, with other calculations for $\mathrm{Im} \, V_{Q\bar{Q}}$ \cite{Albacete:2008dz,Hayata:2012rw}.

\subsection*{Estimating $\Gamma_{Q\bar{Q}}$ for the $\Upsilon(1S)$ state in a strongly coupled $\mathcal{N}=4$ SYM plasma}

We may rewrite the estimate \eqref{eq:thermalwidth2} in a dimensionless form
\begin{equation}
\label{eq:thermalwidthAdS1}
\frac{\Gamma_{Q\bar{Q}}}{T} = -\frac{4}{(Ta_0)^3} \int\limits_0^{\infty} dw \, e^{-\frac{2w}{Ta_0}} w^2 \frac{\mathrm{Im} \, V_{Q\bar{Q}}}{T}(w),
\end{equation}
where $w = LT$. In the case of $\mathcal{N} = 4$ SYM, since $\mathrm{Im} \, V_{Q\bar{Q}}/T$  is only a function of $w=LT$ the only dependence of $\Gamma_{Q\bar{Q}}/T$ on the temperature is via the weight factor $\rho(w) = \exp{(-2w/Ta_0)} w^2$. The position of the ``strip'' in Fig.\ \ref{fig:ImVQQLT} is independent of $T a_0$. Note that as we increase (decrease) $T$, $\rho(w)$ shifts to the right (left, respectively).

We will adopt two approaches to estimate the thermal width. The first one consists of using only the ``strip'' in Fig.\ \ref{fig:ImVQQLT} - this means that we will neglect the region $LT > LT_{\max}$ where our framework does not provide $\mathrm{Im} \, V_{Q\bar{Q}}$. We call this the ``conservative" approach. The second one consists in using the approximation $LT \sim b y_h$ in \eqref{eq:ImVQQAdScase}, ignoring the fact that for $LT \sim LT_{max}$ this approximation ceases to be valid - this will be called the ``extrapolation".\footnote{The authors of \cite{Noronha:2009da} used this second approximation. However, the fact that we must impose $\mathrm{Im} \, V_{Q\bar{Q}} <0$ was not considered - the expression \eqref{eq:ImVQQAdScase} was used (for a fixed $T$) from $L=0$ to $L \to \infty$ instead from $L_{min}$ to $L_{max}$. Excluding from the integration the region $0 < L <L_{min}$ we obtain that the estimate of $\Gamma_{\Upsilon (1S)}$ in \cite{Noronha:2009da} is increased from 48 MeV to 165 MeV.}

\begin{figure}[htp!]
\begin{center}
\includegraphics[width=0.7 \textwidth]{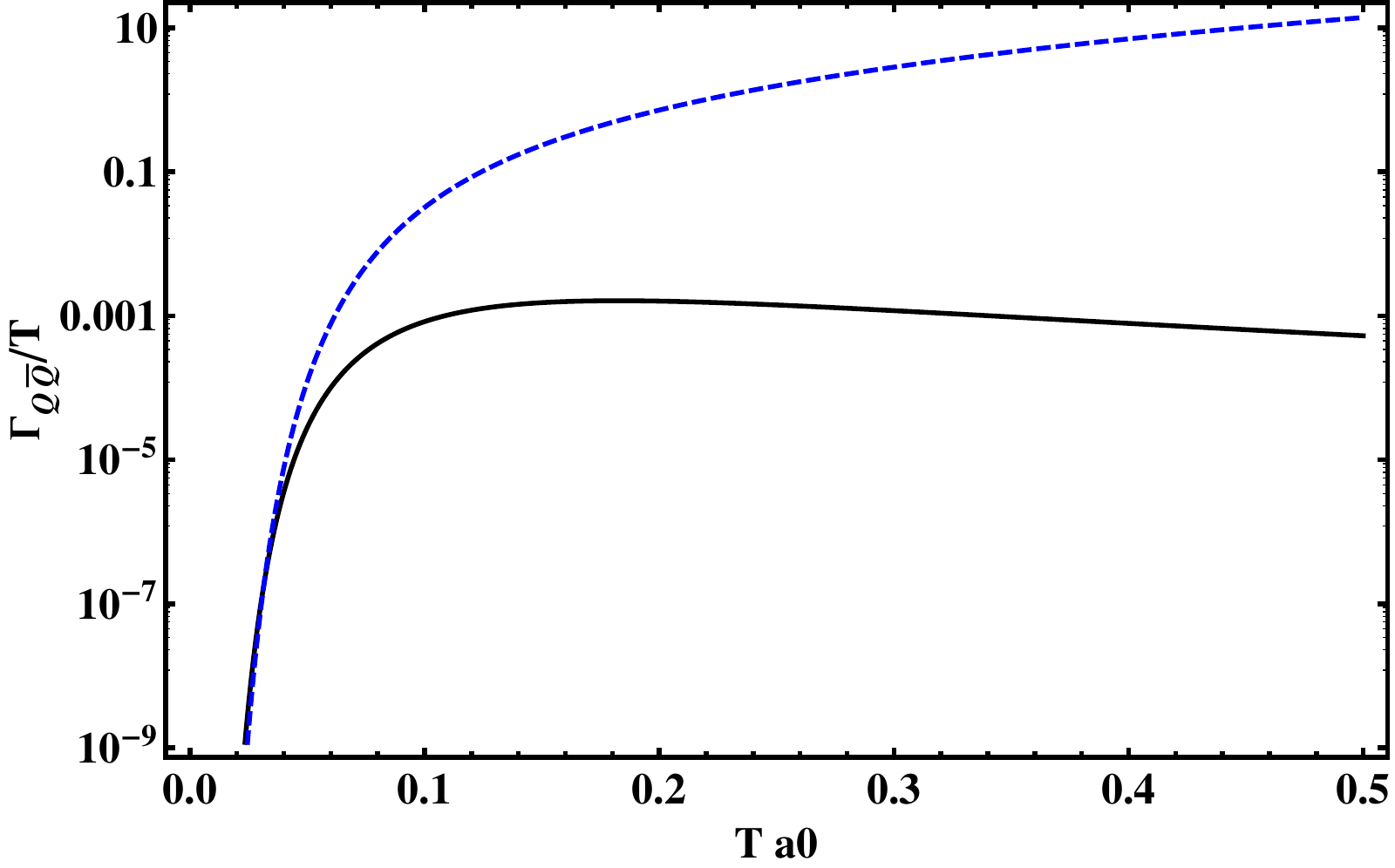}
\caption{The thermal width $\Gamma_{Q\bar{Q}}$ of the $\Upsilon(1S)$ state in $\mathcal{N}=4$ SYM divided by the temperature $T$ as a function of $T a_0$ in a logarithmic scale (the t'Hooft coupling is $\lambda=9$). The solid black curve corresponds to the conservative approach and the dashed blue curve is the extrapolation explained in the text.}
\label{fig:GammaQQ}
\end{center}
\end{figure}

In Fig.\ \ref{fig:GammaQQ} we show $\Gamma_{Q\bar{Q}}/{T}$ for the $\Upsilon(1S)$ state as a function of $T a_0$ for $\lambda=9$. We see that the conservative approach gives a thermal width that can be three orders of magnitude smaller than that computed using the extrapolation. For $a_0 \sim 0.6 \, \mathrm{GeV^{-1}}$, $T \sim 0.5 \, \mathrm{GeV}$, the thermal width varies from 0.5 MeV to 1.5 GeV between the conservative approach and the extrapolation. Therefore, the extrapolation considerably overestimates the thermal width while the conservative approach only gives a lower bound for this quantity. 

\begin{figure}[htp!]
\begin{center}
\includegraphics[width=0.7 \textwidth]{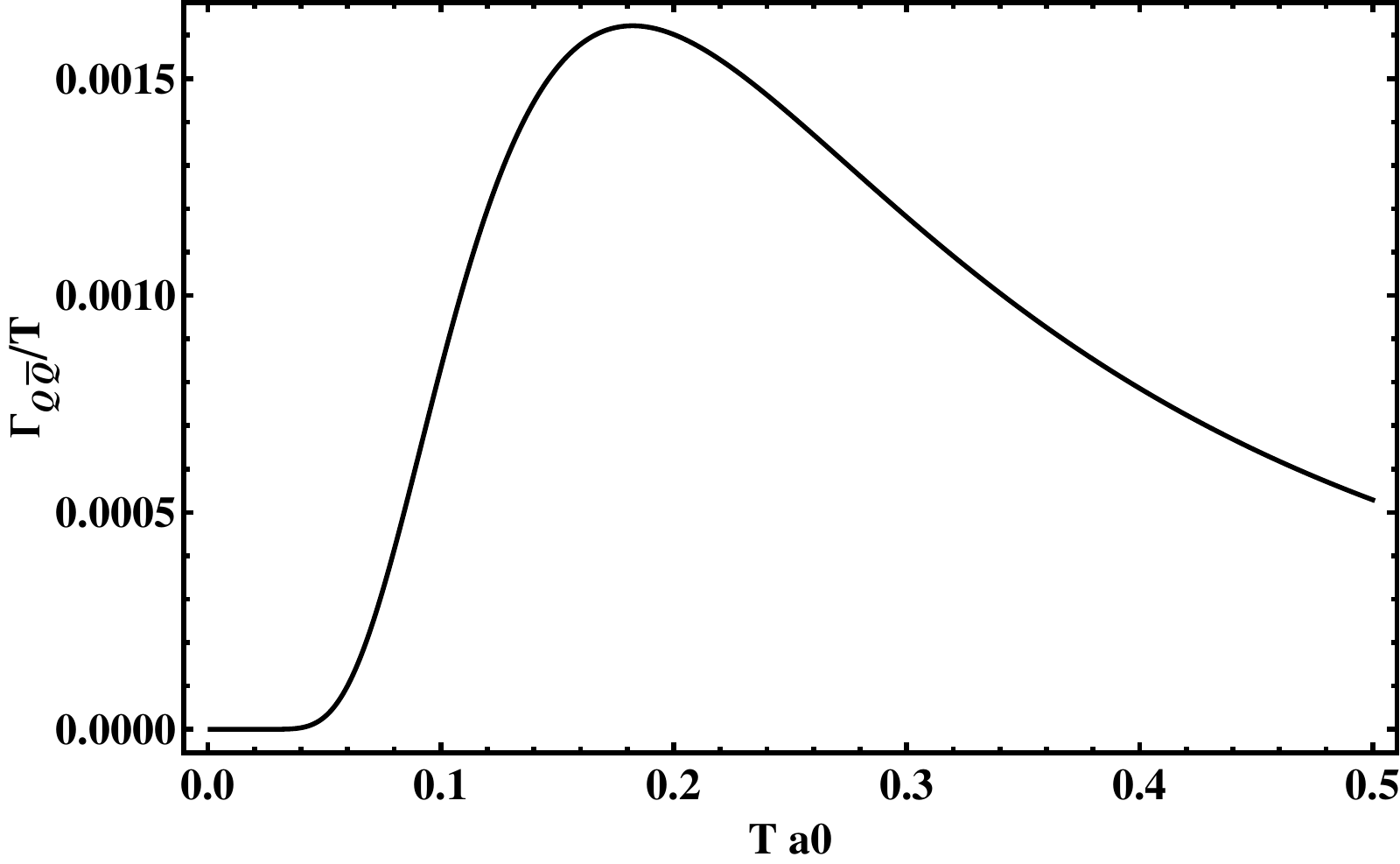}
\caption{The same as in Fig.\ \ref{fig:GammaQQ} but this time the result for the conservative approach is shown in detail.}
\label{fig:GammaQQ2}
\end{center}
\end{figure}

The result for the conservative approach, shown in more detail in Fig.\ \ref{fig:GammaQQ2}, can be understood qualitatively as follows: the weight factor $\rho(w)$ samples only the small region of LT in which $\mathrm{Im} \, V_{Q\bar{Q}} \neq 0$. As one increases the temperature, $\rho(w)$ shifts to the right. For $LT_{min,max} \sim T a_0$ the overlap between $\rho(w)$ and $\mathrm{Im} \, V_{Q\bar{Q}} \neq 0$ happens at the maximum of $\rho(w)$ at $w=Ta_0$ - this corresponds to the maximum in Fig.\ \ref{fig:GammaQQ}. By increasing $T$ even further, the overlap occurs before the maximum of $\rho(w)$ and $\Gamma_{Q\bar{Q}}$ decreases. The temperature dependence of $\Gamma_{Q\bar{Q}}/T$ found in this case is qualitatively similar to that found in recent lattice calculations \cite{Aarts:2011sm}. 

\subsection{Gauss-Bonnet gravity}

\subsection*{Action and metric}

We now consider a class of bulk theories that includes curvature squared corrections to the supergravity action for which the conjectured viscosity bound $\eta/s \geq 1/4\pi$ \cite{Kovtun:2004de} can be violated. The action for these models, called Gauss-Bonnet gravity \cite{Zwiebach:1985uq,Buchel:2008vz,Buchel:2009sk}, is
\begin{align}
\label{eq:gaussbonnetaction}
S=&\frac{1}{16 \pi G_5} \int d^5x \sqrt{-G} \left[ \left( \mathcal{R} + \frac{12}{R^2} \right) + \right. \nonumber \\ & \left. + \frac{\lambda_{GB}}{2} R^2 \left(\mathcal{R}^2-4 \mathcal{R}_{\mu \nu} \mathcal{R}^{\mu \nu} + \mathcal{R}_{\mu \nu \rho \sigma} \mathcal{R}^{\mu \nu \rho \sigma} \right) \right],
\end{align}
where $G_5$ is the five dimensional Newton constant, $\mathcal{R}_{\mu \nu \rho \sigma}$ is the Riemann tensor, $\mathcal{R}_{\mu \nu}$ is the Ricci tensor, $\mathcal{R}$ is the Ricci scalar, and $\lambda_{GB}$ is a constant. The first parenthesis is the usual Einstein-Hilbert + cosmological constant action. The second parenthesis gives the curvature squared corrections. For this particular choice of curvature squared corrections, metric fluctuations in a given background have the same quadratic terms as Einstein gravity. The action \eqref{eq:gaussbonnetaction} has an exact black-brane solution \cite{Cai:2001dz} given by
\begin{equation}
\label{eq:metricgaussbonnet}
ds^2 = - a^2 f_{GB}(U) dt^2 + \frac{U^2}{R^2} d\vec{x}^2 +\frac{dU^2}{f_{GB}(U)},
\end{equation}
where
\begin{equation}
\label{eq:agaussbonnet}
a^2 = \frac{1}{2} \left(1+ \sqrt{1-4\lambda_{GB} } \right) \quad \text{and}
\end{equation}
\begin{equation}
\label{eq:fGBgaussbonnet}
f_{GB} (U) = \frac{U^2}{R^2} \frac{1}{2 \lambda_{GB}} \left[1-\sqrt{1 - 4\lambda_{GB} \left(1-\frac{U_h^4}{U^4} \right) } \right].
\end{equation}
The black brane horizon is the simple root of $f_{GB}(U)$, $U_h$. The plasma temperature is $T = a U_h/(\pi R^2)$. From \eqref{eq:metricgaussbonnet} we see that the AdS radius is given by $aR$ instead of just $R$. In particular, the 't Hooft coupling of the dual strongly coupled CFT is given by $\lambda = a^4 R^4/\alpha'$. The functional form of $a$ and $f_{GB}$ implies that $\lambda_{GB} < 1/4$. However, in practice $\lambda_{GB} \leq 9/100$ to avoid causality violation at the boundary \cite{Brigante:2008gz,Brigante:2007nu}.

The constant $\lambda_{GB}$ is related to the ratio of the shear viscosity $\eta$ and the entropy density $s$ by \cite{Brigante:2008gz,Brigante:2007nu,Kats:2007mq}
\begin{equation}
\label{eq:etasgaussbonnet}
\frac{\eta}{s} = \frac{1}{4\pi} (1- 4 \lambda_{GB}).
\end{equation}
For $\lambda_{GB} > 0$ the viscosity bound for gauge theories with gravity duals, $\eta/s > 1/4\pi$, is violated. The constraint $\lambda_{GB} \leq 9/100$ implies that $\frac{4 \pi \eta}{s} \geq 16/25$.

The evaluation of the real part of the heavy quark potential in the strongly coupled plasma dual to Gauss-Bonnet gravity \eqref{eq:metricgaussbonnet} was already performed in \cite{Noronha:2009ia} (see also \cite{Fadafan:2011gm,AliAkbari:2009pf}). In this section we extend the analysis of \cite{Noronha:2009ia} to include the numerical evaluation of ${\rm Re}\,V_{Q\bar{Q}}$ and also the calculation of the imaginary part of the potential using the worldsheet fluctuation method. Moreover, we give an estimate of the dependence of the Debye screening mass in this theory as a function of $\eta/s$.

\subsection*{Polyakov loop and the real part of the heavy quark potential}

Using the formulas \eqref{eq:equationforL}, \eqref{eq:potentialreg}, and \eqref{eq:polyloopreg} one obtains for the regularized heavy quark free energy
\begin{equation}
\label{eq:polygaussbonnet}
\frac{F_{Q}}{T} = - \frac{R^2}{2 \alpha'} = - \frac{\sqrt{\lambda}}{2a^2}
\end{equation}
while
\begin{equation}
\label{eq:Lgaussbonnet}
LT (y_h) = \frac{2 a}{\pi} y_h \sqrt{2 \bar{f}_{GB} (1,y_h) \lambda_{GB}} \int\limits_1^{\infty} \left[ y^8 \bar{f}_{GB}(1,y_h)^2-y^4 \bar{f}_{GB}(y,y_h) \bar{f}_{GB}(1,y_h) \right]^{-1/2}
\end{equation}
and the real part of the heavy quark potential is given by
\begin{equation}
\label{eq:Fgaussbonnet}
\frac{{\rm Re}\,V_{Q\bar{Q}}}{T} = \frac{R^2}{\alpha'} \frac{1}{y_h} \left\{ \int\limits_1^{\infty}dy\, \left[ \left(1+ \frac{1}{\frac{y^4 \bar{f}_{GB}(y,y_h)}{\bar{f}_{GB}(1,y_h)}} \right)^{1/2} -1 \right] - 1 \right\}
\end{equation}
where $\bar{f}_{GB}(y,y_h)$ is a reduced form of $f_{GB}(U)$ defined by
\begin{equation}
\label{eq:fGBredgaussbonnet}
\bar{f}_{GB}(y,y_h) = 1-\sqrt{1 - 4\lambda_{GB} \left(1-\frac{y_h^4}{y^4} \right) }.
\end{equation}
For $\lambda_{GB}\neq 0$, both \eqref{eq:Lgaussbonnet} and \eqref{eq:Fgaussbonnet} cannot be evaluated in terms of hypergeometric functions. In the limit $LT \ll 1$ one can show \cite{Noronha:2009ia} that
\begin{equation}
\label{eq:Fapproxgaussbonnet}
\frac{{\rm Re}\,V_{Q\bar{Q}}}{T} = -\frac{4 \pi^2 \sqrt{\lambda}}{\Gamma(1/4)^4 LT}\left(1+\frac{c}{a^6\sqrt{1-4\lambda_{GB}}} (LT)^4 \right)\,,
\end{equation}
where $c$ is the constant given by \eqref{eq:constantapprox}\footnote{In \cite{Noronha:2009ia} this corresponds to Eq.\ (34), which can be obtained after some manipulations involving gamma functions. Here we have not performed the entropy subtraction done in \cite{Noronha:2009ia} to obtain their Eq.\ (35).}.

\begin{figure}[htp!]
\begin{center}
\includegraphics[width=0.7 \textwidth]{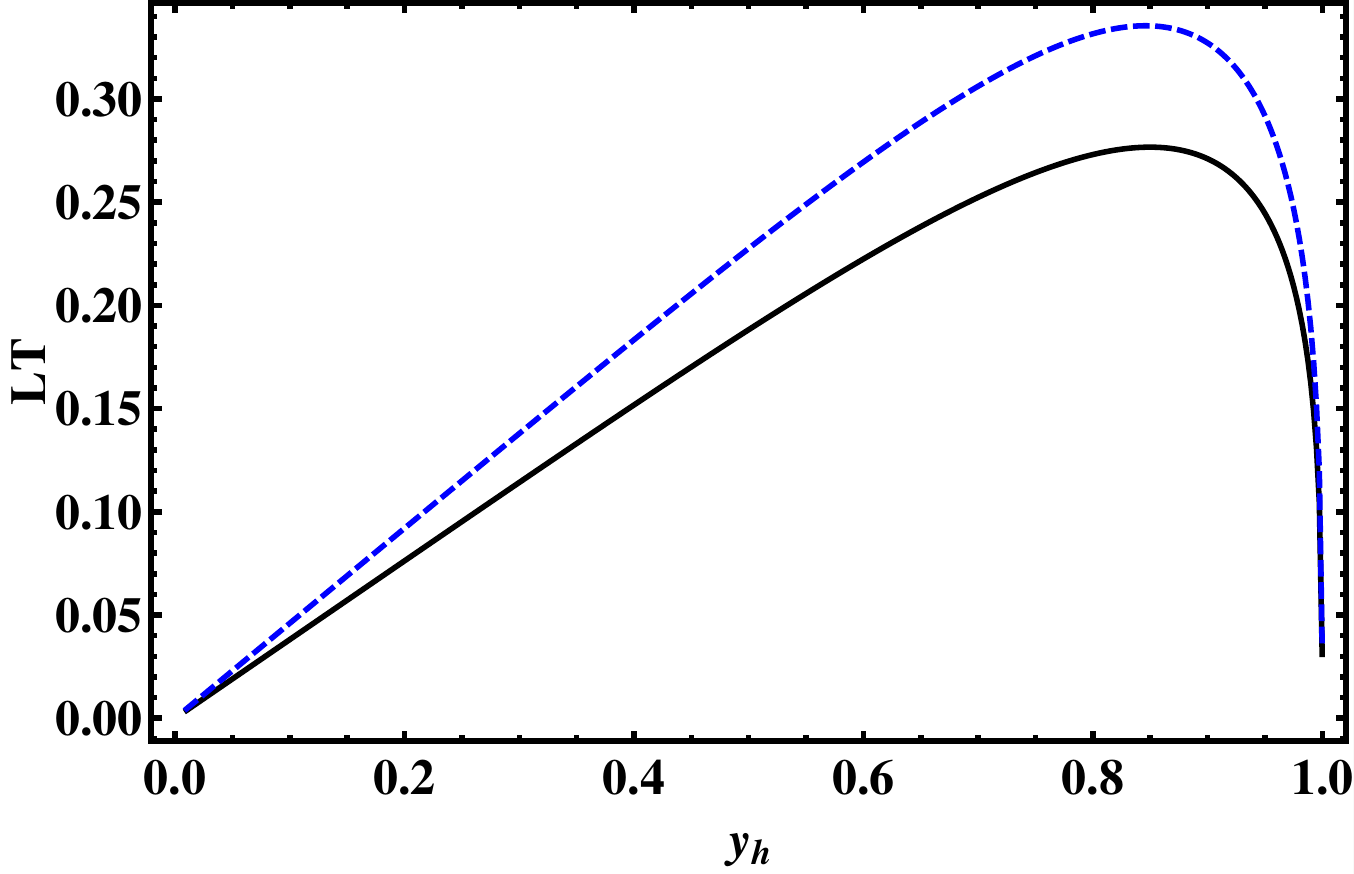}
\caption{$LT$ as a function of $y_h$ in the CFT dual to Gauss-Bonnet gravity. The solid black curve is the result for $\lambda_{GB} = 0$ ($4\pi \eta/s = 1$); the dashed blue curve is the result for $\lambda_{GB} =- 0.25$ ($4\pi \eta/s = 2$).}
\label{fig:gaussbonnetLT}
\end{center}
\end{figure}

\begin{figure}[htp!]
\begin{center}
\includegraphics[width=0.7 \textwidth]{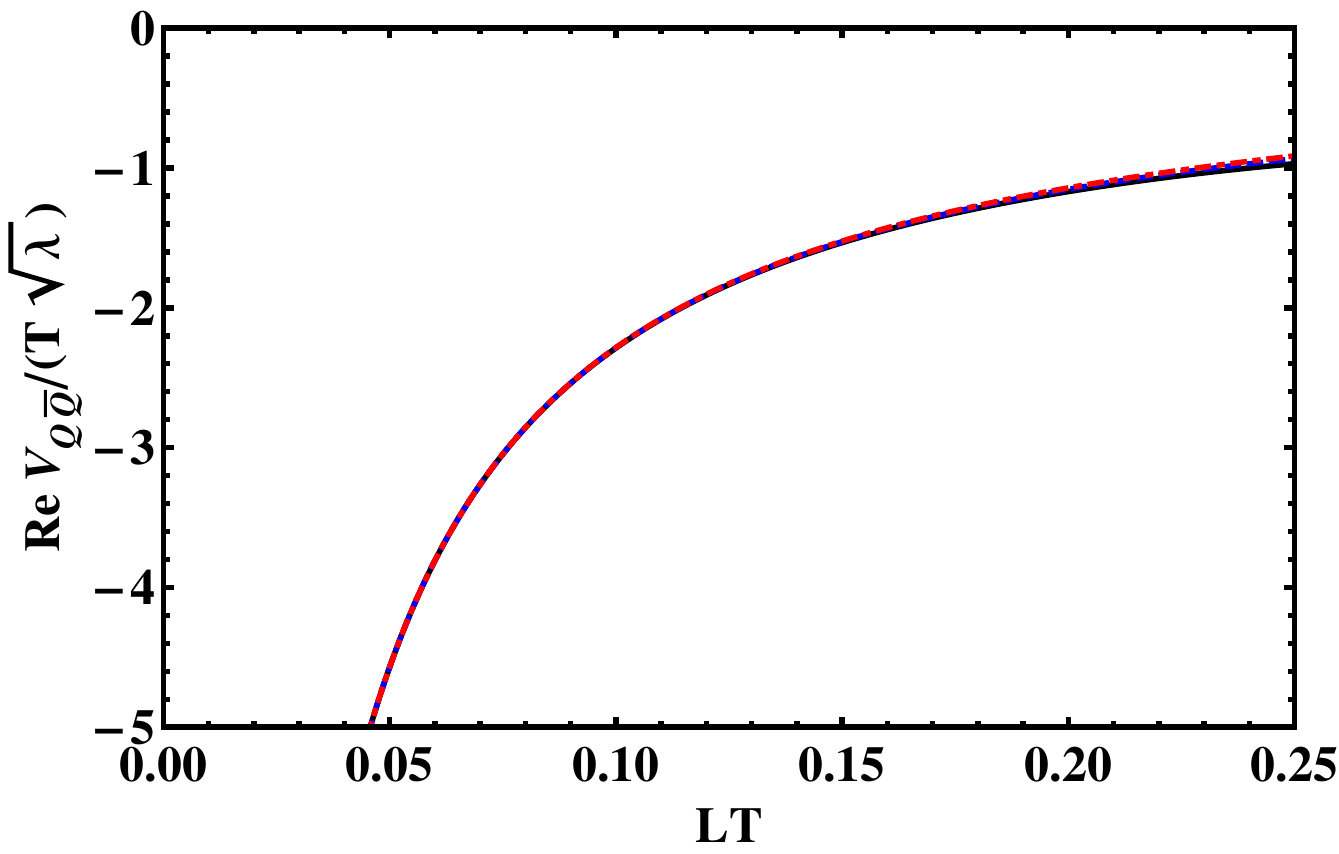}
\caption{${\rm Re}\,V_{Q\bar{Q}}/T$ as a function of $LT$ in the CFT dual to Gauss-Bonnet gravity. The solid black curve is the result for $\lambda_{GB} = 0$ ($4\pi \eta/s = 1$); the dotted-dashed red curve is the result for $\lambda_{GB} = -0.25$ ($4\pi \eta/s = 2$); the dotted blue curve corresponds to the approximation in \eqref{eq:Fapproxgaussbonnet}.}
\label{fig:gaussbonnetFT}
\end{center}
\end{figure}

We can also evaluate \eqref{eq:Lgaussbonnet} and \eqref{eq:Fgaussbonnet} numerically by fixing $\lambda_{GB}$ and using $y_h$ as a parameter. In Fig.\ \ref{fig:gaussbonnetLT} we show $LT$ as a function of $y_h$ for $\lambda_{GB} = 0$ ($4\pi \eta/s = 1$) and $\lambda_{GB} = -0.25$ ($4\pi \eta/s = 2$). We see that increasing $\lambda_{GB}$ (decreasing $\eta/s$) lowers $y_{h,max}$ and $LT_{max}$. However, as shown in Fig.\ \ref{fig:gaussbonnetFT}, the behavior of ${\rm Re}\,V_{Q\bar{Q}}$ as a function of $LT$ does not change significantly with $\lambda_{GB}$. Moreover, one sees that the approximation in \eqref{eq:Fapproxgaussbonnet} is excellent for the values of $\lambda_{GB}$ considered here. In the end, the main effect of increasing $\lambda_{GB}$ is to reduce the allowed interval for $LT$.

\subsection*{Estimate of the Debye mass and its dependence on $\eta/s$}

Using the simple fitting procedure described in section \ref{sec:adsdebyemass} we can obtain a simple estimate for the Debye screening mass in GB gravity and its dependence with $\eta/s$. We use, as before, the model \eqref{eq:kmsmodelF-conf} (with $\sigma=0$). Since we do not have exact expressions for $LT$ and ${\rm Re}\,V_{Q\bar{Q}}/T$ in this case, we cannot prove whether the real part of the potential computed using the classical string shows exponential Debye screening or not. In any case, the cautionary remarks previously made for $\mathcal{N}=4$ SYM are still applicable here and must be kept in mind.

\begin{figure}[htp!]
\begin{center}
\includegraphics[width=0.7 \textwidth]{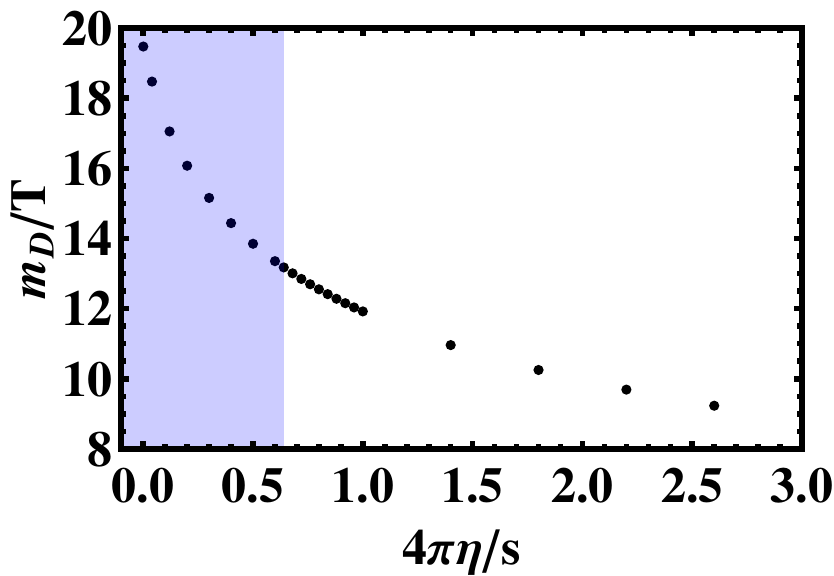}
\caption{An estimate for the Debye screening mass $m_D$ as a function of $4\pi \eta/s$ in the strongly coupled conformal plasma dual to Gauss-Bonnet gravity. The shaded blue region can be excluded since $m_D$ in this region was determined using values of $\lambda_{GB}$ that lead to problems with causality.}
\label{fig:gaussbonnetmD}
\end{center}
\end{figure}

The fitting procedure is done as before for the case of SYM. Varying the values of $\lambda_{GB}$ (therefore, $\eta/s$) we obtain the results for $m_D$ shown in Fig.\ \ref{fig:gaussbonnetmD} (the parameters $\delta$ and $\tilde{C}_1$ do not vary appreciably with respect to those found in the SYM calculation). Here we consider both positive $\lambda_{GB}$ (corresponding to $4 \pi \eta /s <1$) and negative $\lambda_{GB}$ ($4 \pi \eta /s > 1$). In Fig.\ \ref{fig:gaussbonnetmD}, the shaded region denotes the result for $m_D$ computed using values of $\lambda_{GB}$ that lead to problems with causality. One can see that $m_D$ decreases with increasing $\eta /s$ for the allowed values of $\lambda_{GB}$. This result is reasonable since larger $\eta/s$ in general means weaker coupling, which in turns implies that heavy quark pairs are less screened by the medium.

\subsection*{Imaginary part of the heavy quark potential in GB gravity}

\begin{figure}[htp!]
\begin{center}
\includegraphics[width=0.7 \textwidth]{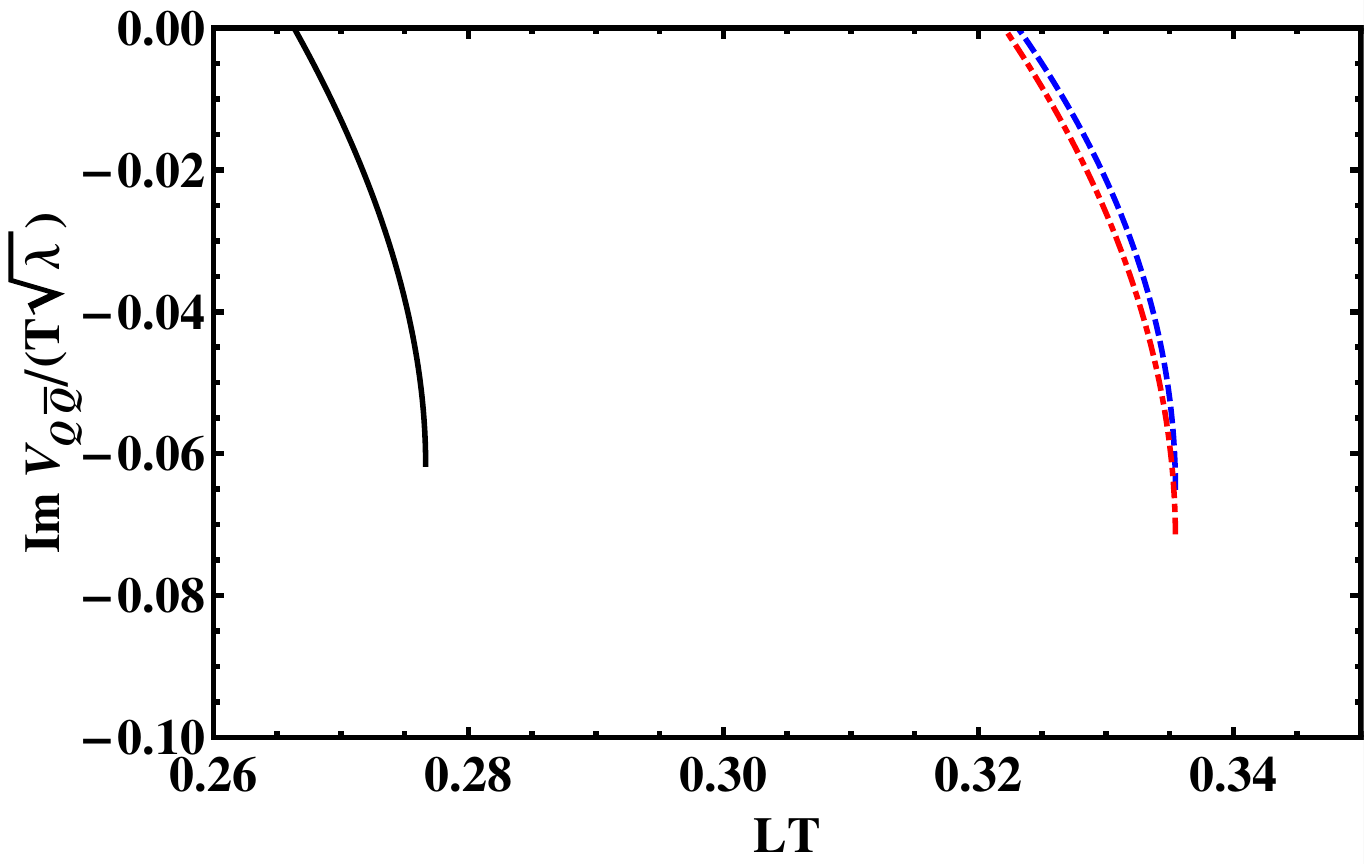}
\caption{$\mathrm{Im} \,V_{Q\bar{Q}}/T$ as a function of $LT$ in the CFT dual to Gauss-Bonnet gravity. The full black curve is the result for $\lambda_{GB} = 0$; the dashed blue curve is the result for $\lambda_{GB} = -0.25$; and the dashed-dotted red curve corresponds to the approximation in Eq.\ \eqref{eq:gaussbonnetImVQQ}.}
\label{fig:gaussbonnetImFQQ}
\end{center}
\end{figure}

Using Eq.\ \eqref{eq:imvqq} we can calculate $\mathrm{Im} \, V_{Q\bar{Q}}$ in this theory and study its dependence on $\eta/s$. The full expression, while easy to derive, is rather cumbersome and therefore omitted in the text. However, a simple expansion for $\lambda_{GB} \ll 1$ results in a more useful expression
\begin{equation}
\label{eq:gaussbonnetImVQQ}
\frac{\mathrm{Im} \,V_{Q\bar{Q}}}{T} = -\frac{\pi \sqrt{\lambda}}{24 \sqrt{2}} \frac{1}{y_h} \left[ (3 y_h^4 - 1) +  \frac{\lambda_{GB}}{3} (9 y_h^4 - 34 y_h^8+ 9 y_h^{12}) \right] + O(\lambda_{GB}^2) .
\end{equation}
For $\lambda_{GB} = 0$ we recover the $\mathcal{N} =4$ SYM result \eqref{eq:ImVQQAdScase}. As before, enforcing $\mathrm{Im}\,V_{Q\bar{Q}}<0$ gives a lower limit for $y_h$ while the condition regarding the validity of the classical string calculations gives a maximum value of $y_h$ (Fig.\ \ref{fig:gaussbonnetLT}). In Fig.\ \ref{fig:gaussbonnetImFQQ} we show the numerical results for $\mathrm{Im} \,V_{Q\bar{Q}}/T$ for $\lambda_{GB} =- 0.25$. Only a small interval of $LT$ is allowed in the conservative approach and increasing $\lambda_{GB}$ shifts this interval to the left. We also see that \eqref{eq:gaussbonnetImVQQ} is a satisfactory approximation to the numerical result for $\lambda_{GB}=-0.25$. 

\subsection{Thermal width of $\Upsilon(1S)$ and its dependence on $\eta/s$}

\begin{figure}[htp!]
\begin{center}
\includegraphics[width=0.7 \textwidth]{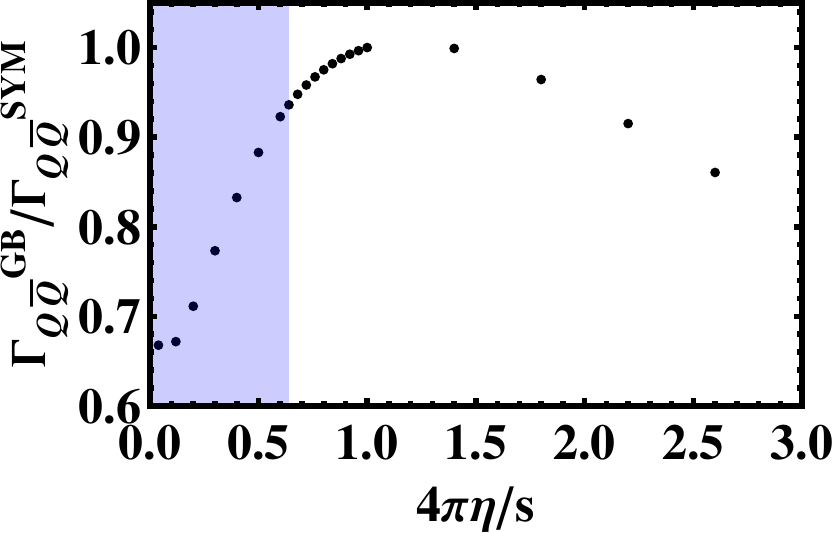}
\caption{Lower bound for $\Upsilon(1S)$ thermal width $\Gamma_{Q\bar{Q}}$ computed via Gauss-Bonnet gravity, normalized by the $\mathcal{N}=4$ SYM result. We used the gauge theory coupling $\lambda = 9$ and $T = 300 \, \mathrm{MeV}$. The shaded blue region denotes the values of the width computed using values of $\lambda_{GB}$ that lead to causality violations in the gauge theory.}
\label{fig:gaussbonnetGQQ}
\end{center}
\end{figure}

In Fig.\ \ref{fig:gaussbonnetGQQ} we present a lower bound for the thermal width $\Gamma_{Q\bar{Q}}$ of the $\Upsilon(1S)$ state as a function of $\eta/s$ for $\lambda = 9$ and $T \sim 300 \, \mathrm{MeV}$. Since changing $\eta/s$ changes the sampling region for $L$, we have again that the shape of Fig.\ \ref{fig:gaussbonnetGQQ} reproduces the shape of the associated ground-state Coulomb wave function. The shaded blue region denotes the values of the width computed using values of $\lambda_{GB}$ that lead to causality violations in the gauge theory. Note that the thermal width, normalized by the value found in strongly coupled SYM, decreases with increasing $\eta/s$. 

\section{Conclusions and Outlook}\label{conclusions}

In this paper we used the gauge/gravity duality to study the imaginary part of the heavy quark potential in strongly coupled plasmas. This imaginary part can be used to estimate the thermal width of heavy quarkonia in strongly coupled plasmas, which may be seen as the strongly coupled analog of the Landau damping induced thermal width found in perturbative QCD calculations \cite{imvrefs}. The thermal worldsheet fluctuation method, originally developed in \cite{Noronha:2009da}, was used here to obtain a lower bound for the thermal width of heavy quarkonium states, such as the $\Upsilon(1S)$, in 2 different holographic toy models of the strongly coupled quark-gluon plasma (QGP): strongly coupled $\mathcal{N}=4$ SYM at large $N_c$ and the strongly coupled CFT dual to GB gravity. Moreover, we proved a general result using the thermal worldsheet fluctuation approach that establishes the connection between the imaginary part of the heavy quark potential at nonzero temperature and the area law of the Wilson loop at zero temperature.  

In the case of strongly coupled SYM we found that the thermal width of $\Upsilon(1S)$ is actually very small in comparison to the plasma temperature for reasonable (and large) values of the t'Hooft coupling. The estimates previously made for this quantity in \cite{Noronha:2009da} have been improved in the present paper and the nontrivial consistency conditions, discussed at length in this manuscript, have conspired to bring down the previous value of the thermal width to values that may be consistent with recent phenomenological models for the quenching of heavy quarkonia in the QGP \cite{Strickland:2011mw,Strickland:2011aa}. It would be interesting to use the imaginary contribution to the heavy quark potential found here to study other quarkonium states \cite{mikenew}. 

Moreover, even though the real part of the heavy quark potential (computed with the classical string approximation) does not show explicit exponential screening in a strongly coupled $\mathcal{N}=4$ SYM plasma, a simple phenomenological estimate for the Debye screening mass can still be extracted via a fit to the real part of the heavy quark potential. Surprisingly enough, this rough estimate for the Debye screening mass is still in fair agreement (within $\sim 11$\%) with the result obtained using the lightest CT-odd supergravity mode \cite{Bak:2007fk}. 

We also computed the thermal width of heavy quarkonia in the CFT dual to GB gravity to study its dependence with $\eta/s$. For a fixed temperature of $T=0.3$ GeV the width has a maximum around $\eta/s=1/4\pi$ and decreases for larger values of $\eta/s$. Following the phenomenological procedure to extract the Debye mass from the real part of the potential described above, we obtained an estimate for the dependence of $m_D$ with $\eta/s$ in this gravity model. Our results suggest that Debye screening effects decrease with increasing $\eta/s$ in a strongly coupled plasma. 

In this paper we assumed that the plasma is isotropic and conformal\footnote{Simple non-conformal models and the respective results for the imaginary part of the potential can be found in Appendix C.} and that the $Q\bar{Q}$ pair is at rest with respect to the thermal bath. It would be interesting to generalize the calculations for the imaginary part of the heavy quark potential performed here by considering gravity models dual to plasmas where these conditions are dropped. For instance, one could compute the thermal width in an anisotropic strongly coupled plasma \cite{Mateos:2011ix,Mateos:2011tv} or in non-conformal gravity models of the QGP such as \cite{Gursoy:2009jd,Gubser:2008ny}.  

Note added: After this paper was submitted to the arxiv we became aware of Refs.\  \cite{Giataganas:2013lga,Fadafan:2013bva} where the imaginary part of the heavy quark potential was computed in a strongly coupled anisotropic plasma using the method described here.

\section*{Acknowledgments}
The authors thank Conselho Nacional de Desenvolvimento Cient\'ifico e Tecnol\'ogico (CNPq) and Funda\c c\~ao de Amparo \`a Pesquisa do Estado de S\~ao Paulo (FAPESP) for support. We also thank A.~Dumitru for discussions about the heavy quark potential and K.~B.~Fadafan for comments on this manuscript. 


\clearpage
\appendix

\section{Covariant expansion of the Nambu-Goto action around the classical solution}\label{covariantmethod}

Expansions of the string action around a given classical solution of the equations of motion, $X_0^\mu(\tau,\sigma)$, are somewhat nontrivial since the worldsheet fluctuations $\delta X^\mu(\tau,\sigma)$ do not transform simply under reparametrization \cite{AlvarezGaume:1981hn}. Thus, the way the fluctuations around the classical solution were included in Section \ref{impartsection}, though correct, are not manifestly covariant. In this section we perform a covariant expansion of the determinant of the worldsheet metric around a generic solution of the classical string equations of motion. 

A fluctuation of the string worldsheet can be written as \cite{AlvarezGaume:1981hn}
\begin{equation}
\label{eq:covfluc}
X_0^{\mu} + \delta X^{\mu} = X_0^{\mu} + \xi^{\mu} -\frac{1}{2} \Gamma^{\mu}_{\rho \lambda}(X_0) \xi^{\rho} \xi^{\lambda} + O(\xi^3)\,
\end{equation}
where $\xi^{\mu}(\tau,\sigma)$ transforms as a vector under reparametrization (which plays the role of Riemann normal coordinates \cite{MTW}). Derivatives with respect to the worldsheet variables $\tau$ and $\sigma$ are given by  
\begin{equation}
\label{eq:XRNC}
\partial_{a} (X_0^{\mu}+\delta X^\mu ) = \partial_a X_0^{\mu} + D_{a} \xi^{\mu} + \frac{1}{3} \mathcal{R}^{\mu}_{\nu \rho \lambda}(X_0) \xi^{\nu} \xi^{\rho} \partial_{a} X_0^{\lambda} + O(\xi^3),
\end{equation}
where $\mathcal{R}^{\mu}_{\nu \rho \lambda}$ is the Riemann curvature tensor and $D_{a}$ is defined as
\begin{equation}
\label{eq:covdev}
D_a \xi^{\mu} \equiv \partial_a \xi^{\mu} + \Gamma^{\mu}_{\nu \rho} (\partial_a X_0^\mu) \xi^{\rho}\,.
\end{equation}
Note that using the chain rule one obtains $D_a \xi^{\nu} = D_{\mu} \xi^{\nu} \partial_a X_0^{\mu}$, where $D_{\mu}$ is the usual space time covariant derivative with an affine connection. This motivates the definition \eqref{eq:covdev} as the covariant derivative of $\xi$ on the worldsheet. The expansion for the background metric becomes
\begin{equation}
\label{eq:RNC}
G_{\mu \nu} (X_0+ \xi) = G_{\mu \nu}(X_0) - \frac{1}{3} \mathcal{R}_{\mu \nu \rho \lambda}(X_0) \xi^{\rho} \xi^{\lambda} + O(\xi^3)\,.
\end{equation}

Using the equations \eqref{eq:XRNC} and \eqref{eq:RNC} we obtain for the induced metric on the worldsheet $h_{ab} = G_{\mu \nu} \partial_{a} X^{\mu} \partial_b X^{\nu}$, up to second order in $\xi$,
\begin{equation}
\label{eq:indmetriccov}
h_{ab} = h^{(0)}_{ab}+ h^{(1)}_{ab} + h^{(2)}_{ab} + O(\xi^3)
\end{equation}
where
\begin{equation}
h^{(0)}_{ab} =\partial_a X_0 \cdot \partial_b X_0,
\end{equation}
\begin{equation}
h^{(1)}_{ab} = \partial_{a} X_0 \cdot D_b \xi + \partial_b X_0 \cdot D_a \xi \quad \quad \mathrm{and}
\end{equation}
\begin{equation}
\label{eq:h2}
h^{(2)}_{ab} =D_a \xi \cdot  D_b \xi + \partial_a X^{\mu}_0 \partial_b X^{\nu}_0 \mathcal{R}_{\mu \nu \rho \lambda} \xi^{\rho} \xi^{\lambda}.
\end{equation}
where the inner product here is defined with respect to the background metric, i.e, $A \cdot B = G_{\mu \nu} A^{\mu} B^{\nu}$. Eq.\ \eqref{eq:indmetriccov} takes into account the effect of worldsheet fluctuations on the induced worldsheet metric in an explicitly reparametrization invariant manner. 

To show that this procedure yields the same results as the non-covariant approach developed in the main text, we use the $AdS_5$-Schwarzschild metric \eqref{eq:metricads} in the formulas above. We also use the static gauge for the worldsheet embedding functions and, thus, $\tau=t$ and $\sigma=x$ and the classical solution is $X_0^\mu=(t,x,0,0,U_c(x))$. As before, the fluctuations are $\delta X^{\mu} = (0,0,0,0, \delta U(x))$. Then, using the inverse of \eqref{eq:covfluc} into \eqref{eq:indmetriccov} and evaluating the induced metric determinant $h = det \,h_{ab}$ we obtain in the end 
\begin{equation}
\label{eq:indmetric}
-h = \left( \frac{dU_c(x)}{dx}\right)^2 + \frac{1}{R^2} (U^4-U_h^4) + \frac{4 U^3}{R^4} \delta U + \frac{6 U^2}{R^4} \delta U ^2 + O(\delta U^3)\,.
\end{equation}
The saddle point approximation for $e^{-S_{NB}}$ can also be obtained by taking the extremum of $h$ with respect to $\delta U$. The extremum of \eqref{eq:indmetric} occurs at $\delta \bar{U} = - U/3$, which yields 
\begin{equation}
\label{eq:metricmin}
-\bar{h} = \left( \frac{dU_c(x)}{dx}\right)^2 + \frac{U^4-3 U_h^4}{3R^4}.
\end{equation}

Now, $dU_c/dx$ is given by the classical solution \eqref{eq:dzdx} and, thus, we obtain the following expression for the (regularized) effective action after integrating over $\tau$ and defining the dimensionless variables $y = U/U_h$ and $y_h=U_h/U_*$
\begin{equation}
\label{eq:actionncov}
S = \frac{\mathcal{T}}{\pi \alpha'} U_*\int^{\infty}_1 dy \left\{\sqrt{\left(\frac{y^4-y_h^4}{y^4-1}\right) - \frac{2}{3} \frac{y^4(1-y_h^4)}{(y^4-1)(y^4-y_h^4)}}-1 \right\}-\frac{\mathcal{T}}{\pi \alpha'} U_*.
\end{equation}

Eq.\ \eqref{eq:actionncov} gives both the real and imaginary parts of $V_{Q\bar{Q}}$. Note that second term inside the square root above represents the contribution from worldsheet fluctuations and this term only becomes relevant close to the bottom of the classical string solution at $U_*$ (also, see that this term is well behaved in the UV, $y\to \infty$, which is expected since it comes solely from thermal effects). The shift in ${\rm Re} V_{Q\bar{Q}}$ due to fluctuations is easier to obtain in the covariant approach and it can be determined from Eq.\ \eqref{eq:actionncov}. For $T=0$ (i.e., $y_h = 0$), \eqref{eq:actionncov} can be evaluated in terms of hypergeometric functions as explained in Appendix \ref{sec:integrals}. The result is
\begin{equation}
\label{eq:potT0cov}
V_{Q\bar{Q}} = -\frac{4 \pi^2}{\Gamma(1/4)^4} \frac{R^2}{\alpha'} \, {}_2 F_1 \left[-\frac{1}{2},-\frac{1}{4} ; \frac{1}{4}; \frac{2}{3} \right] \frac{1}{L}\,.
\end{equation}
Since ${}_2 F_1 [-1/2,-1/4 ; 1/4; 2/3 ] = 1.38$, we see that long wavelength worldsheet fluctuations change the vacuum result for $\mathcal{N}=4$ SYM by $\sim 40 \%$ (which can be accommodated, for instance, by rescaling the t'Hooft coupling).

\begin{figure}[htp!]
\begin{center}
\includegraphics[width=0.7 \textwidth]{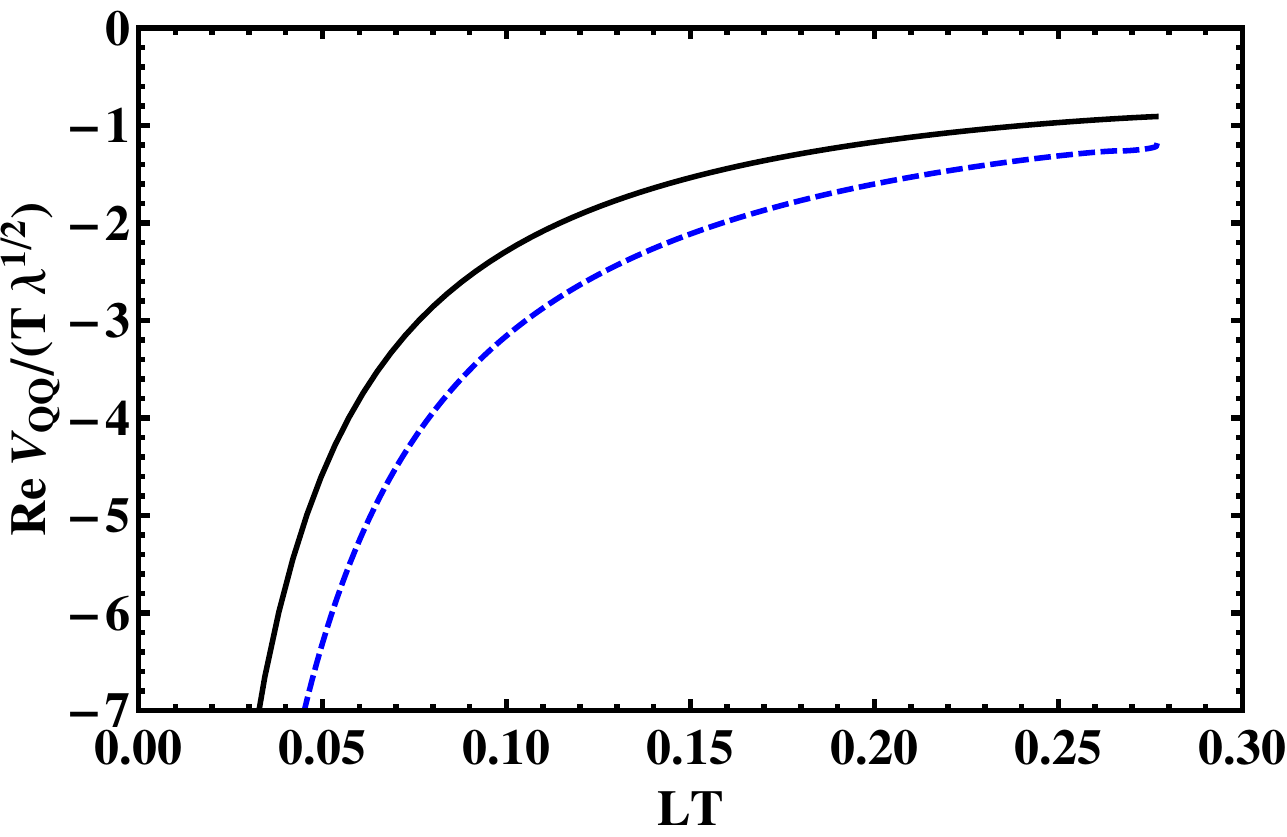}
\caption{$\mathrm{Re}\, V_{Q\bar{Q}}/(T\sqrt{\lambda})$ as a function of $LT$ for the strongly coupled $\mathcal{N}=4$ SYM plasma. The solid line is the real part calculated without considering thermal fluctuations on the string worldsheet while the dashed line is the real part of the potential including the fluctuations.}
\label{fig:reVQQcov}
\end{center}
\end{figure}

\begin{figure}[htp!]
\begin{center}
\includegraphics[width=0.7 \textwidth]{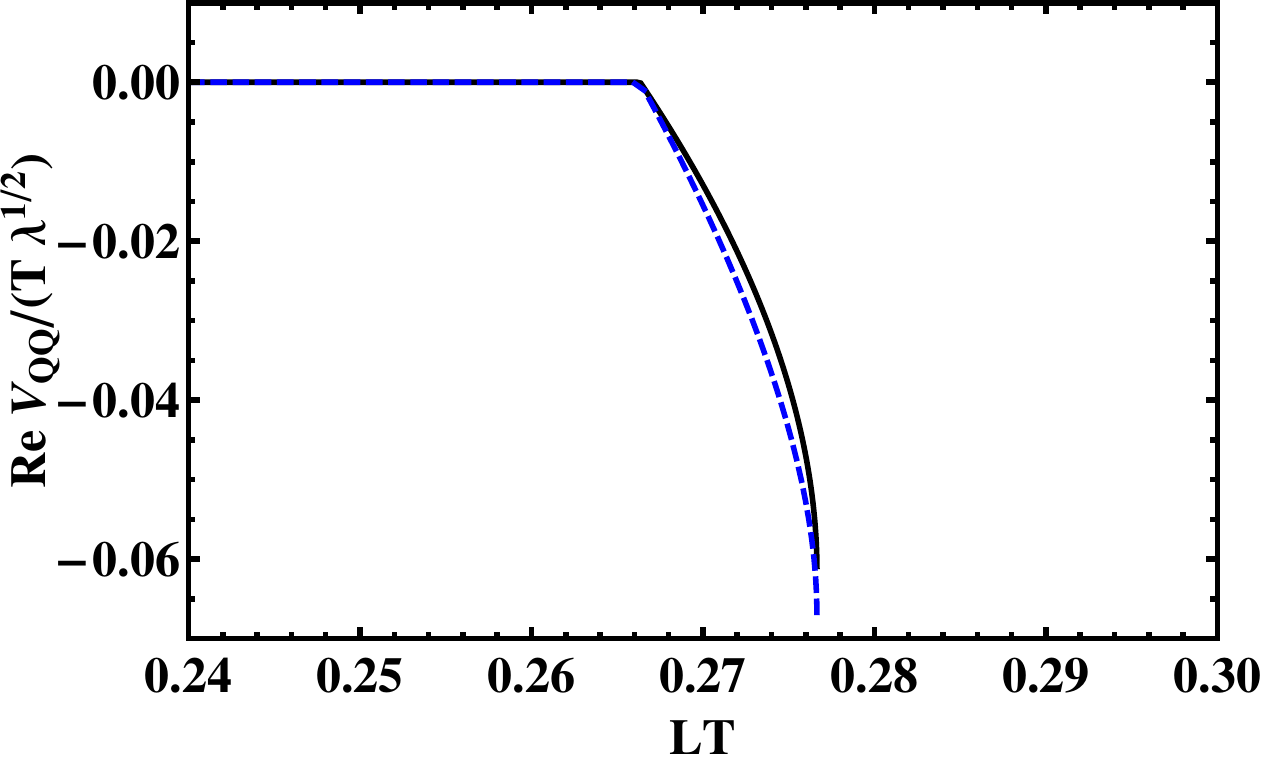}
\caption{$\mathrm{Im}\, V_{Q\bar{Q}}/(T\sqrt{\lambda})$ as a function of $LT$ for the strongly coupled $\mathcal{N}=4$ SYM plasma. The solid (dashed) line is the result from the non-covariant (covariant) method, respectively.}
\label{fig:imVQQcov}
\end{center}
\end{figure}

In Fig.\ \ref{fig:reVQQcov} we show the effect of fluctuations on the real part of the potential while in Fig.\ \ref{fig:imVQQcov} we compare the results for the imaginary part of the potential computed using the covariant method and the non-covariant method developed in the main text. The real part of the part of the potential changes slightly due to fluctuations while the imaginary part is almost unaffected by the choice of method. This is expected since in the non-covariant approach we focus mainly on fluctuations near the bottom of the string while in the covariant approach long-wavelength fluctuations along the whole worldsheet are taken into account. Since the imaginary part is generated by the fluctuations near the bottom of the string both approaches are equivalent to determine $\mathrm{Im} \, V_{Q\bar{Q}}$.

\newpage

\section{Some useful formulas for the evaluation of Wilson loops}\label{sec:integrals}

In this Appendix we present some useful techniques to evaluate the integrals found in the calculation of Wilson loops via the gauge/gravity correspondence. All the integrals and properties studied in this section can be found, for example, in \cite{gradshteyn}. The main idea is to use integral representations of the beta and the (Gaussian) hypergeometric functions to perform the integrals that appear in the study of holographically computed Wilson loops.

\subsection*{Beta function}

A recurring integral found in these calculations is the beta function
\begin{equation}
\label{eq:betafunction}
\mathrm{B} (a,b) \equiv \int_0^1 t^{a-1} (1-t)^{b-1} dt
\end{equation}
with $\mathrm{Re} (a)$, $\mathrm{Re} (b)>0$. This function satisfies the reflection property
\begin{equation}
\label{eq:betafunction-reflection}
\mathrm{B} (a,b) = \mathrm{B}(b,a)
\end{equation}
and is related to the gamma function by
\begin{equation}
\label{eq:betafunction-gamma}
\mathrm{B}(a,b) = \frac{\Gamma(a) \Gamma(b)}{\Gamma(a+b)}\,.
\end{equation}

For example, for $\mathcal{N} = 4$ SYM at strong 't Hooft coupling (and $T=0$) one finds that the relation between $U_*$ and $L$ is given by Eq.\ \eqref{eq:LAdST0}
\begin{equation}
\label{eq:LAdST0-app1}
\frac{L}{2} = \frac{R^2}{U_*} \int \limits_{1}^{\infty} dy \, \frac{1}{y^2\sqrt{y^4-1}}.
\end{equation}
Therefore, changing variables to $t = 1/y^4$ and using \eqref{eq:betafunction} one finds
\begin{equation}
\label{eq:LAdST0-app2}
\frac{L}{2} = \frac{R^2}{U_*} \frac{1}{4} \mathrm{B}(3/4,1/2) = \frac{R^2}{U_*} \frac{ \Gamma(1/2) \Gamma(3/4)}{\Gamma(5/4)}.
\end{equation}

To simplify \eqref{eq:LAdST0-app2} a bit further it is useful to remember that the gamma function satisfies
\begin{equation}
\label{eq:gamma-factorial}
\Gamma(z+1) = z \Gamma(z) \quad \text{and}
\end{equation}
\begin{equation}
\label{eq:gamma-factorial}
\Gamma(1-z) \Gamma(z) = \frac{\pi}{\sin(\pi z)}.
\end{equation}
Therefore, $\Gamma(1/2) = \sqrt{\pi}$. Moreover, we have that  $\Gamma(3/4) = \sqrt{2} \pi/\Gamma(1/4)$ and $\Gamma(5/4) = \Gamma(1/4)/4$. We finally obtain \eqref{eq:LU0AdST0}
\begin{equation}
\label{eq:LAdST0-app3}
\frac{L}{2} = \frac{R^2}{U_*} \frac{\sqrt{2} \pi^{3/2}}{\Gamma(1/4)^2}.
\end{equation}
 
The same procedure can be applied to $V_{Q\bar{Q}}$ in Eq.\ \eqref{eq:VQQregAdST0}. To avoid having $a$ or $b$ with  a negative real part, one introduces a factor $y^{\gamma}$ in the integrand, performs the integration, and then takes $\gamma \to 0$. This gives the expression in Eq.\ \eqref{eq:VQQU0AdST0}.

\subsection*{Gaussian hypergeometric function}

The Gaussian hypergeometric function ${}_2 F_1$ can be defined by the power series
\begin{align}
\label{eq:hypergeofunction}
{}_2 F_1(a,b;c;z) \equiv & 1+ \frac{ab}{1!\, c} z + \frac{a(a+1)b(b+1)}{2!\,c(c+1)} z^2 + \nonumber \\ & + \cdots + \frac{a(a+1)\cdots(a+n)b(b+1)\cdots(b+n)}{n! \, c(c+1)\cdots(c+n)} z^n + \cdots,
\end{align}
where $a,b,c,z$ are real numbers, with $c \neq -1,-2,\cdots$. The series converges for $|z|<1$ while for the rest of the complex plane ${}_2 F_1$ is obtained by analytic continuation.

We are mainly interested in the following integral representation of ${}_2 F_1$
\begin{equation}
\label{eq:hypergeofunction-integral}
{}_2 F_1(a,b;c;z) = \frac{1}{\rm B(b-c,b)} \int_0^1 t^{b-1} (1-t)^{c-b-1} (1-zt)^{-a} dt\,.
\end{equation}
This representation is valid for $|z|<1$ and for $\mathrm{Re} (c) > \mathrm{Re} (b)>0$. This relation follows immediately  from the binomial theorem and Eq.\ \eqref{eq:betafunction}.

Eq.\ \eqref{eq:hypergeofunction-integral} was used in \eqref{eq:LTAdSTneq0} for thermal $\mathcal{N} = 4$ SYM at strong 't Hooft coupling to find
\begin{equation}
\label{eq:LTAdSTneq0-app1}
LT(y_h) = \frac{2}{\pi} y_h \sqrt{1-y_h^4} \int\limits^{\infty}_1 \frac{dy}{\sqrt{(y^4-y_h^4)(y^4-1)}}.
\end{equation}
Applying the change of variables $t = 1/y^4$ we find that
\begin{equation}
\label{eq:LTAdSTneq0-app2}
LT(y_h) = \frac{2 \sqrt{2\pi}}{\Gamma(1/4)^2} y_h \sqrt{1-y_h^4}\, {}_2 F_1\left[\frac{1}{2},\frac{3}{4};\frac{5}{4};y_h^4 \right].
\end{equation}
The same procedure can be applied to determine ${\rm Re}\,V_{Q\bar{Q}}/T$ in Eq.\ \eqref{eq:FQQAdSTneq0}, which leads to  \eqref{eq:FQQAdSTneq0-res}
\begin{equation}
\label{eq:FQQAdSTneq0-app2}
\frac{{\rm Re}\,V_{Q\bar{Q}}}{T} = - \frac{R^2} {\alpha'} \frac{\sqrt{2\pi^3}}{\Gamma(1/4)^2} \frac{1}{y_h} \, {}_2 F_1 \left[-\frac{1}{2},-\frac{1}{4} ; \frac{1}{4};y_h^{4} \right].
\end{equation}

The series definition of ${}_2 F_1$ \eqref{eq:hypergeofunction} also simplifies the derivation of the series expansion in \eqref{eq:FQQAdSTneq0approx}. For example, for $LT < 1$ we find, up to linear terms on $y_h^4$,
\begin{equation}
\label{eq:approxapp1}
\frac{LT}{2}\frac{\Gamma(1/4)^2}{\sqrt{2 \pi} y_h} = 1- \frac{1}{5} y_h^4.
\end{equation}
For ${\rm Re}\,V_{Q\bar{Q}}/T$ in \eqref{eq:FQQAdSTneq0-app2} we obtain
\begin{equation}
\label{eq:approxapp2}
-\frac{{\rm Re}\,V_{Q\bar{Q}}}{T} \frac{\alpha'}{R^2} \frac{\Gamma(1/4)^2}{\sqrt{2} \pi^{3/2}} y_h = 1+\frac{1}{2} y_h^4.
\end{equation}
Therefore, after multiplying \eqref{eq:approxapp1} and \eqref{eq:approxapp2} and using \eqref{eq:approxapp1} to zeroth order in $y_h^4$ we obtain the following expression valid to order $(LT)^4$ \eqref{eq:FQQAdSTneq0approx}
\begin{equation}
\label{eq:approxapp3}
\frac{{\rm Re}\,V_{Q\bar{Q}}}{T} = -\frac{4 \pi^2 \sqrt{\lambda}}{\Gamma(1/4)^4 LT}(1+c(LT)^4),
\end{equation}
with $c=3 \Gamma(1/8)^8/(5 \cdot 2^7 \pi^2)$. The same reasoning also shows that the series expansion of ${\rm Re}\,V_{Q\bar{Q}}/T$ is of the form $\left[1+a_4(LT)^4+a_8(LT)^8+\cdots \right]/(LT)$.

As a last remark, note that the derivative of ${}_2 F_1$ with respect to $z$ is 
\begin{equation}
\label{eq:hypergeofunction-derivative}
\frac{d}{dz} {}_2 F_1(a,b;c;z) = \frac{ab}{c} {}_2 F_1(a+1,b+1;c+1;z)\,.
\end{equation}
If the real part of the potential has the general form $\mathcal{V} \propto e^{-m_D/T(LT)}/(LT)^\delta$ then
\begin{equation}
\label{eq:notexpscreen-app}
\frac{d}{d(LT)} \left[ (LT)^\delta\frac{\mathcal{V}}{T} \right] = -\frac{m_D}{T} (LT)^\delta \frac{\mathcal{V}}{T}\,.
\end{equation}
However, by \eqref{eq:hypergeofunction-derivative}, the holographically computed potential given by \eqref{eq:LTAdSTneq0-app2} and \eqref{eq:FQQAdSTneq0-app2} does not satisfy this condition because
\begin{equation}
\label{eq:notexpscreen-app}
\frac{d}{d(LT)} \left[ (LT)^\delta\frac{{\rm Re}\,V_{Q\bar{Q}}}{T} \right] = \frac{d \left[(LT)^\delta\frac{{\rm Re}\,V_{Q\bar{Q}}}{T} \right] / dy_h^4}{d(LT)/dy_h^4 } \neq -\frac{m_D}{T} (LT)^\delta \frac{{\rm Re}\,V_{Q\bar{Q}}}{T}\,.
\end{equation}

\newpage

\section{Some other results involving Wilson loops}
\label{sec:dpstacks}

In this section we apply the formalism developed in the main text to calculate heavy quark potentials and their imaginary parts in a slightly more general class of gravity duals which include, as an interesting subset, the low energy theories of coincident stacks of Type II Dp-branes. While some of these results were initially discussed in \cite{Brandhuber:1998er}, as far as we know, a complete evaluation of $V_{Q\bar{Q}}$ and its imaginary part have not been presented before.

This section is organized as follows. First we present the class of metrics we use. We then calculate the Polyakov loop and ${\rm Re}\,V_{Q\bar{Q}}$. An approximation for small $L$ is discussed. Finally, we show the results for the imaginary part of $V_{Q\bar{Q}}$ in these theories.

\subsection{Gravity duals considered}

We consider the gravity duals described by the following metric (in the string frame)\footnote{In principle, we could generalize this metric a bit further by making the change $2 \alpha \to \gamma$ in the exponent of the terms inside the square brackets. However, the expressions obtained cannot be integrated using the method discussed in \ref{sec:integrals}. Moreover, the analysis of the UV divergence gets more involved since in this case the metric is not asymptotic $AdS$. For these reasons, we keep the form of the metric shown in \eqref{eq:metricdadslike}.}
\begin{equation}
\label{eq:metricdadslike}
ds^2 = \left( \frac{U}{R} \right)^{\alpha} \left\{ -\left[1-\left(\frac{U_h}{U}\right)^{2 \alpha} \right] dt^2 + dx_i dx^i \right\} + \left( \frac{R}{U} \right)^{\beta} \left[1-\left(\frac{U_h}{U}\right)^{2 \alpha} \right]^{-1} dU^2
\end{equation}
where $R$ is a constant, $i$ runs from 1 to $D-1$ and $D$ is the total number of dimensions of the corresponding gauge theory. From the confinement criteria \cite{Kinar:1998vq}, we see that as long as $\alpha \geq \beta$ the theory does not confine (in the sense of an area law for the rectangular Wilson loop).

The black brane temperature is
\begin{equation}
\label{eq:tempadslike}
T = \frac{\alpha}{2\pi R} \left( \frac{U_h}{R} \right)^{\frac{\alpha+\beta-2}{2}}
\end{equation}
and the entropy density is
\begin{equation}
\label{eq:entropyadslike}
s = \frac{1}{4G_5} \left( \frac{U_h}{R} \right)^{\frac{3\alpha}{2}}.
\end{equation}

\subsection*{Polyakov loop}

We start by calculating the Polyakov loop in this class of theories. The unregularized expression for the heavy quark free energy is given by 
\begin{equation}
\label{eq:polyakovloopadslikenreg}
F^{nreg}_Q = \frac{1}{2\pi \alpha'} R^{\frac{\beta-\alpha}{2}} U_*^{\frac{2+\alpha-\beta}{2}} \int_{y_h}^{\infty} y^{\frac{\alpha-\beta}{2}} dy.
\end{equation}
We have three possibilities. If $\alpha-\beta < -2$, then there is no UV divergence. If $\alpha-\beta = -2$, the integral diverges logarithmically. If $\alpha - \beta > -2$ the UV divergence is worse than logarithmic. If $\alpha-\beta \geq -2$ we use the temperature independent regularization $\int_0^{\infty} dy \, y^{(\alpha-\beta)/2}$ and, with this choice, the final regularized expression for $F_Q$ is the same regardless of the sign of $\alpha-\beta$
\begin{equation}
\label{eq:polyakovloopadslikereg}
F_Q = -\frac{1}{\pi \alpha'} \frac{R^{\frac{\beta-\alpha}{2}} U_*^{\frac{\alpha-\beta+2}{2}}}{2+\alpha-\beta} y_h^{\frac{\alpha-\beta+2}{2}} \quad \text{if} \quad \alpha-\beta \geq -2
\end{equation}
and the Polyakov loop is simply $|\langle{\rm tr\,}\bold{L}(T) \rangle |=e^{-F_Q/T}$. 

\subsection*{Real part of the heavy quark potential}

We can now proceed to the calculation of the real part of the heavy quark potential. Using \eqref{eq:actionnotreg} and adopting the regularization used for $F_Q$, we have
\begin{equation}
\label{eq:Ladslikereg}
\frac{L}{2} = \frac{R^{\frac{\alpha+\beta}{2}}}{U_*^{\frac{\alpha+\beta-2}{2}}} \sqrt{1-y_h^{2\alpha}} \int\limits_1^{\infty} dy \, \frac{y^{\frac{\alpha-\beta}{2}}}{\sqrt{(y^{2\alpha}-y_h^{2\alpha})(y^{2\alpha}-1)}}
\end{equation} 
and
\begin{equation}
\label{eq:FQQadslikereg}
{\rm Re}\,V_{Q\bar{Q}} = \frac{R^{\frac{\beta-\alpha}{2}} U_*^{\frac{2+\alpha-\beta}{2}}}{\pi \alpha'} \left\{ \left[ \int\limits_1^{\infty} dy \, y^{\frac{\alpha-\beta}{2}} \left( \sqrt{\frac{y^{2\alpha}-y_h^{2\alpha}}{y^{2\alpha}-1}} -1 \right) \right] -\frac{2}{\alpha-\beta+2} \right\}\,.
\end{equation}

The evaluation of the integrals in both \eqref{eq:Ladslikereg} and \eqref{eq:FQQadslikereg} proceed as discussed before. The results are
\begin{equation}
\label{eq:wilsonloopadslikereg2}
\frac{L}{2} = \frac{R^{\frac{\alpha+\beta}{2}}}{U_*^{\frac{\alpha+\beta-2}{2}}} \sqrt{1-y_h^{2\alpha}} \frac{1}{2\alpha} \mathrm{B} \left( \frac{5\alpha-\beta-2}{4\alpha},\frac{1}{2}\right) {}_2 F_1 \left[\frac{1}{2},\frac{5\alpha-\beta-2}{4\alpha} ; \frac{7\alpha-\beta-2}{4\alpha};y_h^{2 \alpha} \right]
\end{equation}
and
\begin{equation}
\label{eq:FQQadslikereg2}
{\rm Re}\,V_{Q\bar{Q}} = \frac{R^{\frac{\beta-\alpha}{2}} U_*^{\frac{2+\alpha-\beta}{2}}}{\pi \alpha'} \frac{1}{2\alpha} \mathrm{B} \left( \frac{\beta-\alpha-2}{4\alpha},\frac{1}{2}\right) {}_2 F_1 \left[-\frac{1}{2},\frac{\beta-\alpha-2}{4\alpha} ; \frac{\alpha+\beta-2}{4\alpha};y_h^{2 \alpha} \right]\,.
\end{equation}

\subsection*{Imaginary part of the heavy quark potential}

We can also calculate $\mathrm{Im} \,V_{Q\bar{Q}}$ via equation \eqref{eq:imvqq} and obtain
\begin{equation}
\label{eq:ImFQQadslike}
\mathrm{Im}\, V_{Q\bar{Q}} =  -\frac{1}{4\sqrt{2} \alpha'} \frac{1}{\alpha} \left( \frac{U_h}{R} \right)^{\frac{\alpha-\beta}{2}} \frac{U_h}{y_h^{\frac{\alpha-\beta+2}{2}}} \frac{1}{4\alpha-2} \left[ (4\alpha-2) y_h^{2\alpha} -2\alpha + 2 \right].
\end{equation}
The condition $\mathrm{Im}\, V_{Q\bar{Q}} < 0 $ implies that
\begin{equation}
\label{eq:ImFQQadslikecond}
y_h > \left( \frac{\alpha-1}{2\alpha-1} \right)^{\frac{1}{2\alpha}}\,.
\end{equation}
Note that one must require that $\alpha > 1/2$ for \eqref{eq:ImFQQadslikecond} to be well defined. One can check that the formulas above give the correct $AdS_5$ limit given by $\alpha = \beta = 2$.

\subsection{Expansion for small $y_h$}

The expressions for $L$ and ${\rm Re}\,V_{Q\bar{Q}}$ in \eqref{eq:wilsonloopadslikereg2} and \eqref{eq:FQQadslikereg2} can be expanded for small $L U_h^{\frac{\alpha+\beta-2}{2}} \sim LT$. This amounts to an expansion in small $y_h$. By the same procedure applied before we obtain in this approximation
\begin{equation}
\label{eq:FQQadslikeapprox}
{\rm Re}\,V_{Q\bar{Q}} \propto \frac{1}{L^{\frac{\alpha-\beta+2}{\alpha+\beta-2}}} \left( 1 + c \, U_h^{2\alpha} L^{\frac{4\alpha}{\alpha+\beta-2}} \right), 
\end{equation}
where $c$ is a positive constant. The gauge theory has conformal behavior (i.e., $V_{Q\bar{Q}} \propto (1/L)(1+c(LT)^{\delta})$ only when $\alpha=\beta=2$, which corresponds to the gravity dual in $AdS_5$.

\subsection{Results for $Dp$-branes}

The results of the previous sections can be applied to a special class of metrics corresponding to the (near horizon) supergravity solutions of stacks of $Dp$-branes in type II superstring theories. We start by writing the supergravity metric (in the string frame) for $N$ coincident near-extremal black $Dp$-branes in the near-horizon limit \cite{Itzhaki:1998dd},
\begin{align}
\label{eq:adspmetric}
ds^2 = & \left(\frac{U}{R}\right)^{\left(\frac{7-p}{2} \right)} \left[ -f(U) dt^2 + dx_i dx^i \right] + \left(\frac{R}{U}\right)^{\left(\frac{7-p}{2} \right)} \frac{dU^2}{f(U)} + \nonumber \\ & +g_{YM} \sqrt{d_p N} U^{\frac{p-3}{2}} d\Omega^2_{8-p} 
\end{align}
where $i$ runs from 1 to $p$,
\begin{equation}
\label{eq:adspradius}
R^{\frac{7-p}{2}} = g_{YM} \sqrt{d_p N},
\end{equation}
\begin{equation}
\label{eq:adspdp}
d_p =  \Gamma \left(\frac{9-p}{2}\right) \frac{2^{11-2p} \pi^{\frac{13-2p}{2}}}{9-p} \quad \text{and}
\end{equation}
\begin{equation}
\label{eq:adspfU}
f(U) = 1-\left(\frac{U_h}{U}\right)^{7-p}.
\end{equation}
The dilaton field $\phi$ is given by
\begin{equation}
\label{eq:adspdilaton}
e^{\phi} = (2\pi)^{2-p} g_{YM}^2 \left(\frac{R}{U} \right)^{\frac{(7-p)(3-p)}{4}}.
\end{equation}

Note that taking $p=3$ in \eqref{eq:adspmetric} corresponds to the $AdS_5$ case. Only in this case the geometry separates in a product of a $p+2$ dimensional spacetime and an $8-p$ dimensional sphere. In the following we assume a fixed configuration for the compact coordinates. Also, note that if $p\neq 3$ the dilaton runs and, thus, the dual gauge theory is not conformal even in the vacuum. 

The metric is now of the form \eqref{eq:metricdadslike} with $\alpha=\beta = (7-p)/2$. The results of the previous sections then apply and the (regularized) heavy quark free energy is
\begin{equation}
\label{eq:adsppolyakov}
F_Q = -\frac{1}{2\pi \alpha'} U_h
\end{equation}
while
\begin{equation}
\label{eq:adspLwilson}
\frac{L}{2} = \frac{R^{7-p}}{U_*^{\frac{5-p}{2}}} \sqrt{1-y_h^{7-p}} \frac{1}{7-p} \mathrm{B} \left( \frac{6-p}{7-p},\frac{1}{2}\right) {}_2 F_1 \left[\frac{1}{2},\frac{6-p}{7-p} ; \frac{19-3p}{14-2p};y_h^{7-p} \right]
\end{equation}
and the real part of the potential is
\begin{equation}
\label{eq:adspFQQwilson}
{\rm Re}\,V_{Q\bar{Q}} =\frac{U_*} {\pi \alpha'} \frac{1}{7-p} \mathrm{B} \left( -\frac{1}{7-p},\frac{1}{2}\right) {}_2 F_1 \left[-\frac{1}{2},-\frac{1}{7-p} ; \frac{5-p}{14-2p};y_h^{7 -p} \right]\,.
\end{equation}
Moreover, one can use  \eqref{eq:imvqq} to find
\begin{equation}
\label{eq:adspImFQQ}
\mathrm{Im} \,V_{Q\bar{Q}} =  -\frac{1}{4\sqrt{2} \alpha'} \frac{1}{(6-p)(7-p)} \frac{U_h}{y_h} \left[ (12-2p) y_h^{7-p} - 5+p \right]\,.
\end{equation}
For the last equation to be valid the following condition must be satisfied
\begin{equation}
\label{eq:ImFQQadslikecond2}
y_h > \left( \frac{5-p}{12-2p} \right)^{\frac{1}{7-p}}\,.
\end{equation}

\newpage

\section{Curvature scalar on the string worldsheet}
\label{sec:ricci}

In this appendix we study the curvature scalar $\mathcal{R}$ associated with the induced metric on the string worldsheet. As a specific example, we will focus on the Schwarzschild/$AdS_5$ metric \eqref{eq:metricads}. Our main aim is to evaluate the curvature scalar at the bottom of the string at finite $LT$, $\mathcal{R}(LT)$, and compare it with the corresponding $T=0$ result, $\mathcal{R}(0)$. If $\mathcal{R}(LT_{max}) \gg \mathcal{R}(0)$, this signals that near the maximum of $LT$, $LT_{max}$, highly curved string worldsheet configurations start to become relevant. This, in particular, means that one should take care in interpreting $LT_{max}$ as a screening length of the quark-antiquark pair.

For the metric \eqref{eq:metricads}, the induced metric $h_{ab} = G_{\mu \nu} \partial_{a} X^{\mu} \partial_b X^{\nu}$ on the string worldsheet configuration for the rectangular Wilson loop (in the static gauge) is given by
\begin{align}
\label{eq:indmetric}
h_{\tau \tau} & = \frac{1}{4 R^2} \left( \frac{U_h^4}{U(x)^2} - U(x)^2 \right), \nonumber \\
h_{\sigma \sigma} & = \frac{1}{4 R^2} \left( U(x)^2 + \frac{4 R^2 U(x)^2 U'(x)^2}{U(x)^4-U_h^4}\right) \mathrm{and} \nonumber \\
h_{\tau \sigma} & = h_{\sigma \tau} = 0.
\end{align}
Computing the curvature scalar $\mathcal{R}$ using this metric and using the equation of motion \eqref{eq:dzdx} to remove $U'(x)$ and $U''(x)$ from the resulting expressions, one finds
\begin{equation}
\label{eq:curvature}
\mathcal{R} = \frac{2 R^6 \left(\frac{R^8 \left(3 U_h^4-U(x)^4\right) \left(U_*^4-U(x)^4\right)^2}{\left(U_h^4-U_*^4\right)^2}+2 U(x)^4 \left(U_h^8-U(x)^8\right)+\frac{\left(U_h^4-U(x)^4\right) \left(6 U_h^4 U(x)^4+U(x)^8-3 U_h^8\right) \left(U(x)^4-U_*^4\right)}{U_h^4-U_*^4}\right)}{U(x)^4 \left(U_h^4-U(x)^4\right)^2 \left(\frac{R^8 \left(U(x)^4-U_*^4\right)}{\left(U_h^4-U_*^4\right) \left(U_h^4-U(x)^4\right)}+U(x)^4-U_h^4\right)^2}.
\end{equation}
At the bottom of the string, $U(0) = U_*$. Then, \eqref{eq:curvature} reduces to ($y_h = U_h/U_*$),
\begin{equation}
\label{eq:curvbottom}
\mathcal{R}(y_h) = - \frac{4 R^6 \left(y_h^4+1\right)}{U_*^8 \left(1-y_h^4\right)^3}.
\end{equation}
The $T=0$ curvature scalar is found by fixing $y_h=0$ in the equation above. In this case, we may use \eqref{eq:LU0AdST0} to obtain $\mathcal{R}$ explicitly as a function of $L$ and obtain
\begin{equation}
\label{eq:curvmalda}
\mathcal{R} (T=0) = - \frac{\Gamma \left(\frac{1}{4}\right)^{16}}{1024 \pi ^{12}} \frac{L^8}{R^{10}}.
\end{equation}
One can see that the curvature scalar is well behaved for any finite $L$ when $T=0$.

\begin{figure}[htp!]
\begin{center}
\includegraphics[width=0.7 \textwidth]{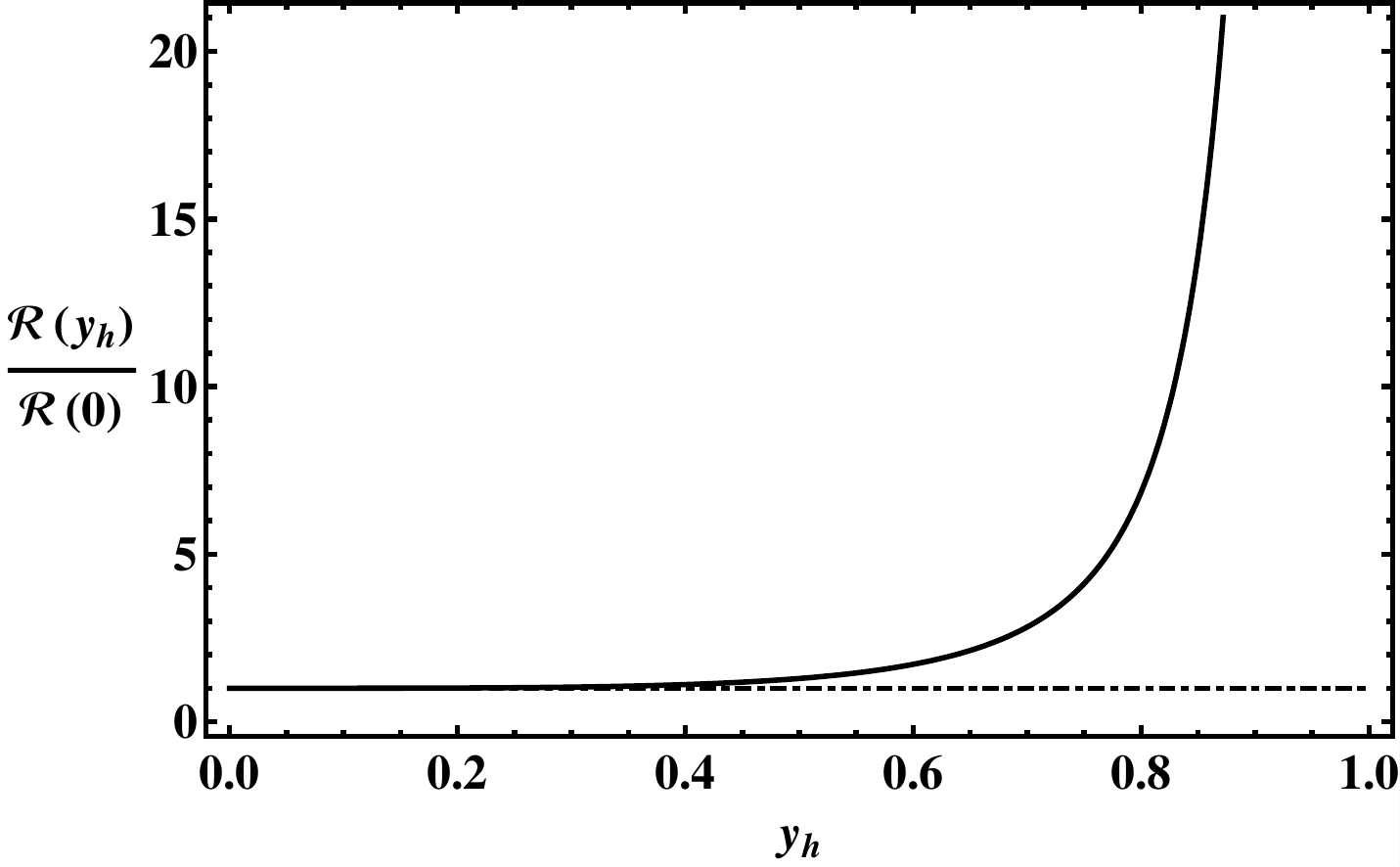}
\caption{The ratio of curvature scalars $\mathcal{R}(y_h)/\mathcal{R}(0)$ as a function of $y_h$ associated with the worldsheet metric for the strongly coupled $\mathcal{N}=4$ SYM plasma. The ratio diverges when the bottom of the string reaches the horizon (where $y_h=1$).}
\label{fig:ricciyh}
\end{center}
\end{figure}

The ratio between the curvature scalars for $T\neq0$ and $T=0$ at the bottom of the string is given by
\begin{equation}
\label{eq:curvratioyh}
\frac{\mathcal{R}(y_h)}{\mathcal{R}(0)} = \frac{1+y_h^4}{(1-y_h^4)^3}.
\end{equation}
Note that this ratio diverges for $y_h \to 1$. This means that a string worldsheet that stretches up to the horizon is highly curved and must receive quantum corrections. In other words, the classical configurations with $y_h > y_{h,max} = 0.85$ are highly curved and must be dealt with care. Already for $y_h = y_{h,max}$, we have $\mathcal{R}(y_{h,max}) \sim 15 \mathcal{R}(0)$. In Fig.\ \ref{fig:ricciyh} we present a plot of the ratio $\mathcal{R}(y_h)/\mathcal{R}(0)$ as a function of $y_h$.

\begin{figure}[htp!]
\begin{center}
\includegraphics[width=0.7 \textwidth]{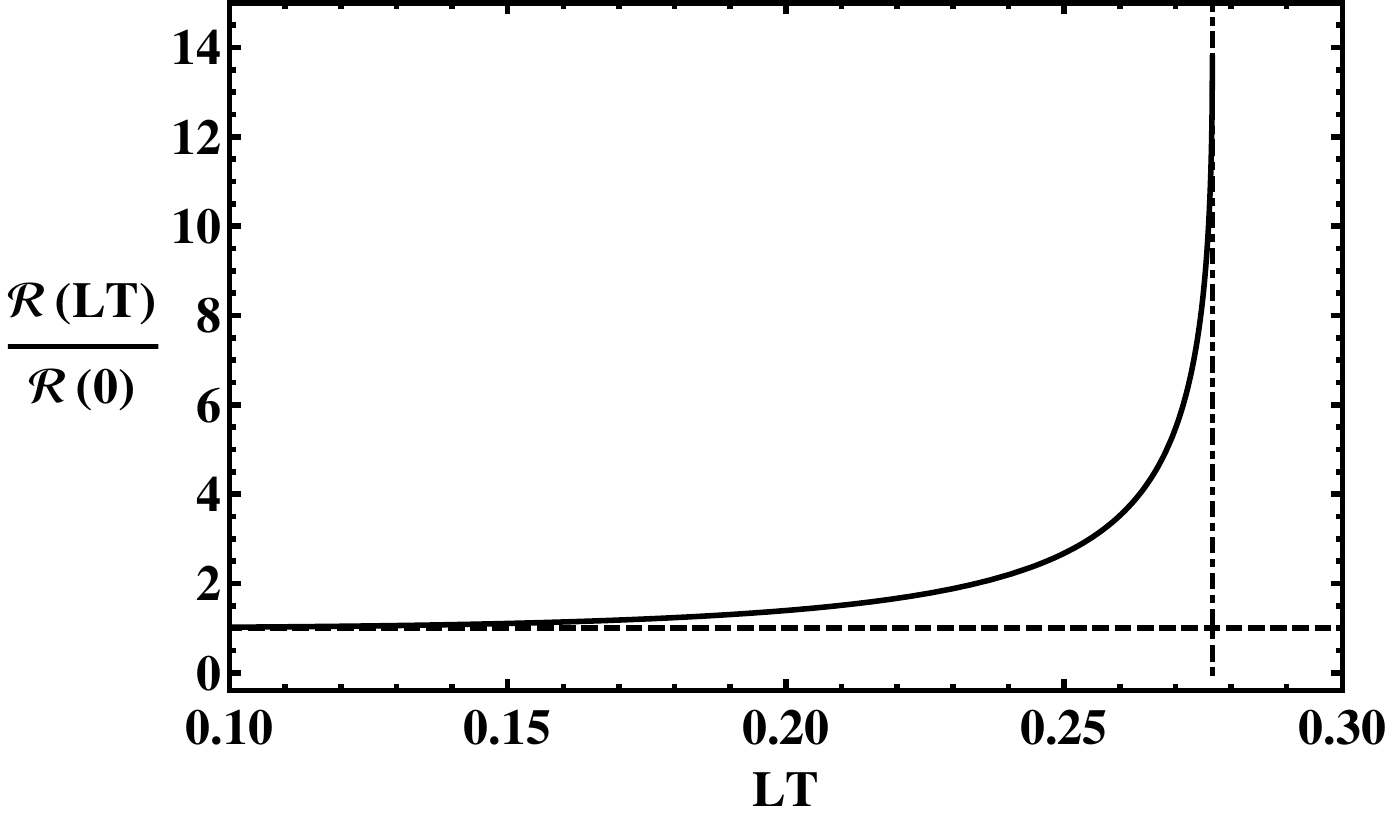}
\caption{The ratio of curvature scalars $\mathcal{R}(LT)\mathcal{R}(0)$ as a function of $LT$, up $LT_{max}$, associated with the worldsheet metric for the strongly coupled $\mathcal{N}=4$ SYM plasma. The vertical line denotes $LT = LT_{max}$ where $\mathcal{R} (LT_{max}) \sim 10 \,\mathcal{R} (0)$.}
\label{fig:ricciLT}
\end{center}
\end{figure}

Now we can use \eqref{eq:LTAdSTneq0-res} to solve for $y_h$ as a function of $LT$ in the branch $0 < y < y_h$ and evaluate $\mathcal{R}$ as a function of $LT$, up to $LT_{max}$, as in Fig.\ \ref{fig:ricciLT}. We see that for $LT \sim LT_{max}$, $\mathcal{R} (LT) \sim 10 \,\mathcal{R} (0)$, which corresponds to a situation of high curvature on the string worldsheet.

\begin{thebibliography}{99}


\bibitem{Wilson:1974sk} 
  K.~G.~Wilson,
  Phys.\ Rev.\ D {\bf 10}, 2445 (1974).
  
\bibitem{wilsonloop}
  J.~-L.~Gervais and A.~Neveu, Nucl.\ Phys.\ B {\bf 163}, 189 (1980); A.~M.~Polyakov, Nucl.\ Phys.\ B {\bf 164}, 171 (1980).
  
\bibitem{kapusta} J.~I.~Kapusta, C.~Gale, {\it Finite Temperature Field Theory, Principles and Applications}, Cambridge University Press, second edition (2006).


\bibitem{polyakov} A.~M.~Polyakov, Phys.\ Lett.\ B {\bf 72}, 477 (1978); G.~'t Hooft, Nucl.\ Phys.\ B {\bf 138}, 1 (1978); {\bf 153}, 141 (1979); B.~Svetitsky and L.~G.~Yaffe, Nucl.\ Phys.\ B {\bf 210}, 423 (1982).





\bibitem{McLerran:1980pk}
  L.~D.~McLerran and B.~Svetitsky,
  Phys.\ Lett.\  B {\bf 98}, 195 (1981);  
  Phys.\ Rev.\  D {\bf 24}, 450 (1981).

  
\bibitem{Kaczmarek:2002mc} 
  O.~Kaczmarek, F.~Karsch, P.~Petreczky, F.~Zantow,
  Phys.\ Lett.\ B {\bf 543}, 41 (2002)
  [hep-lat/0207002].
  
\bibitem{Philipsen:2010gj} 
  O.~Philipsen,
  arXiv:1009.4089 [hep-lat].
  
\bibitem{imvrefs} M.~Laine et al., JHEP {\bf 0703} (2007) 054; JHEP {\bf 0705}, 028
(2007); A.~Beraudo, J.~P.~Blaizot and C.~Ratti, Nucl.\ Phys.\ A {\bf 806}, 312 (2008); N.~Brambilla, J.~Ghiglieri, A.~Vairo, P.~Petreczky, Phys.
Rev. D {\bf 78}, 014017 (2008).

\bibitem{otherrefs1} C.~Miao, A.~Mocsy, and P.~Petreczky, Nucl.\ Phys.\
A {\bf 855}, 125 (2011), arXiv:1012.4433 [hep-ph]; N.~Brambilla, M.~A.~Escobedo, J.~Ghiglieri, J.~Soto, and A.~Vairo,
JHEP {\bf 09}, 038 (2010), arXiv:1007.4156 [hep-ph].

\bibitem{otherrefsImV} Y.~Burnier, M.~Laine, and M.~Vepsalainen, (2009),
arXiv:0903.3467 [hep-ph]; A.~Dumitru, Y.~Guo,
and M.~Strickland, Phys.\ Rev.\ D {\bf 79}, 114003 (2009),
arXiv:0903.4703 [hep-ph]; O.~Philipsen and M.~Tassler,
(2009), arXiv:0908.1746 [hep-ph].

\bibitem{Rothkopf:2011db} 
  A.~Rothkopf, T.~Hatsuda, S.~Sasaki,
  Phys.\ Rev.\ Lett.\  {\bf 108}, 162001 (2012)
  [arXiv:1108.1579 [hep-lat]].
  
\bibitem{Noronha:2009da} 
  J.~Noronha and A.~Dumitru,
  Phys.\ Rev.\ Lett.\  {\bf 103}, 152304 (2009)
  [arXiv:0907.3062 [hep-ph]].
  
\bibitem{gaugegravityduality} 
  J.~M.~Maldacena,
  Adv.\ Theor.\ Math.\ Phys.\  {\bf 2}, 231 (1998)
  [Int.\ J.\ Theor.\ Phys.\  {\bf 38}, 1113 (1999)]
  [hep-th/9711200]; E.~Witten, Adv.\ Theor.\ Math.\ Phys.\ {\bf 2}, 253 (1998); {\bf 2}, 505 (1998); S.~S.~Gubser,
I.~R.~Klebanov and A.~M.~Polyakov, Phys.\ Lett.\ B {\bf 428}, 105 (1998).


\bibitem{Maldacena:1998im} 
  J.~M.~Maldacena,
  Phys.\ Rev.\ Lett.\  {\bf 80}, 4859 (1998)
  [hep-th/9803002].

\bibitem{Brandhuber:1998bs} 
  A.~Brandhuber, N.~Itzhaki, J.~Sonnenschein and S.~Yankielowicz,
  Phys.\ Lett.\ B {\bf 434}, 36 (1998)
  [hep-th/9803137].
 
\bibitem{Rey:1998bq} 
  S.~-J.~Rey, S.~Theisen and J.~-T.~Yee,
  Nucl.\ Phys.\ B {\bf 527}, 171 (1998)
  [hep-th/9803135].

\bibitem{Bak:2007fk} 
  D.~Bak, A.~Karch and L.~G.~Yaffe,
  JHEP {\bf 0708}, 049 (2007)
  [arXiv:0705.0994 [hep-th]].
  
\bibitem{kiritsisbook} E.~Kiritsis, ``String Theory in a Nutshell", Princeton University Press, 2007.

\bibitem{Noronha:2009ud} 
  J.~Noronha,
  Phys.\ Rev.\ D {\bf 81}, 045011 (2010)
  [arXiv:0910.1261 [hep-th]].
  
\bibitem{Noronha:2010hb} 
  J.~Noronha,
  Phys.\ Rev.\ D {\bf 82}, 065016 (2010)
  [arXiv:1003.0914 [hep-th]].

\bibitem{Grigoryan:2011cn} 
  H.~R.~Grigoryan and Y.~V.~Kovchegov,
  Nucl.\ Phys.\ B {\bf 852}, 1 (2011)
  [arXiv:1105.2300 [hep-th]].



\bibitem{Kinar:1998vq} 
  Y.~Kinar, E.~Schreiber and J.~Sonnenschein,
  Nucl.\ Phys.\ B {\bf 566}, 103 (2000)
  [hep-th/9811192].

\bibitem{Sonnenschein:1999if} 
  J.~Sonnenschein,
  hep-th/0003032.
  
\bibitem{Albacete:2008dz} 
  J.~L.~Albacete, Y.~V.~Kovchegov and A.~Taliotis,
  Phys.\ Rev.\ D {\bf 78}, 115007 (2008)
  [arXiv:0807.4747 [hep-th]].

\bibitem{Hayata:2012rw} 
  T.~Hayata, K.~Nawa and T.~Hatsuda,
  arXiv:1211.4942 [hep-ph].



\bibitem{Zwiebach:1985uq} 
  B.~Zwiebach,
  Phys.\ Lett.\ B {\bf 156}, 315 (1985).
  
\bibitem{Buchel:2008vz} 
  A.~Buchel, R.~C.~Myers and A.~Sinha,
  JHEP {\bf 0903}, 084 (2009)
  [arXiv:0812.2521 [hep-th]].
  
\bibitem{Buchel:2009sk} 
  A.~Buchel, J.~Escobedo, R.~C.~Myers, M.~F.~Paulos, A.~Sinha and M.~Smolkin,
  JHEP {\bf 1003}, 111 (2010)
  [arXiv:0911.4257 [hep-th]].

\bibitem{Noronha:2009vz} 
  J.~Noronha, M.~Gyulassy and G.~Torrieri,
  arXiv:0906.4099 [hep-ph]; 
  J.~Noronha, M.~Gyulassy and G.~Torrieri,
  Phys.\ Rev.\ C {\bf 82}, 054903 (2010)
  [arXiv:1009.2286 [nucl-th]].
  




\bibitem{Karsch:1987pv} 
  F.~Karsch, M.~T.~Mehr and H.~Satz,
  Z.\ Phys.\ C {\bf 37}, 617 (1988).


\bibitem{Aarts:2011sm} 
  G.~Aarts, C.~Allton, S.~Kim, M.~P.~Lombardo, M.~B.~Oktay, S.~M.~Ryan, D.~K.~Sinclair and J.~I.~Skullerud,
  JHEP {\bf 1111}, 103 (2011)
  [arXiv:1109.4496 [hep-lat]].
  
\bibitem{Kovtun:2004de} 
  P.~Kovtun, D.~T.~Son and A.~O.~Starinets,
  Phys.\ Rev.\ Lett.\  {\bf 94}, 111601 (2005)
  [hep-th/0405231].

\bibitem{Cai:2001dz} 
  R.~-G.~Cai,
  Phys.\ Rev.\ D {\bf 65}, 084014 (2002)
  [hep-th/0109133].
  
\bibitem{Brigante:2008gz} 
  M.~Brigante, H.~Liu, R.~C.~Myers, S.~Shenker and S.~Yaida,
  Phys.\ Rev.\ Lett.\  {\bf 100}, 191601 (2008)
  [arXiv:0802.3318 [hep-th]].

\bibitem{Brigante:2007nu} 
  M.~Brigante, H.~Liu, R.~C.~Myers, S.~Shenker and S.~Yaida,
  Phys.\ Rev.\ D {\bf 77}, 126006 (2008)
  [arXiv:0712.0805 [hep-th]].
  
\bibitem{Kats:2007mq} 
  Y.~Kats and P.~Petrov,
  JHEP {\bf 0901}, 044 (2009)
  [arXiv:0712.0743 [hep-th]].

\bibitem{Noronha:2009ia} 
  J.~Noronha and A.~Dumitru,
  Phys.\ Rev.\ D {\bf 80}, 014007 (2009)
  [arXiv:0903.2804 [hep-ph]].
  
\bibitem{Fadafan:2011gm} 
  K.~B.~Fadafan,
  Eur.\ Phys.\ J.\ C {\bf 71}, 1799 (2011)
  [arXiv:1102.2289 [hep-th]].
  
\bibitem{AliAkbari:2009pf} 
  M.~Ali-Akbari and K.~Bitaghsir Fadafan,
  Nucl.\ Phys.\ B {\bf 835}, 221 (2010)
  [arXiv:0908.3921 [hep-th]].
  
\bibitem{Strickland:2011mw} 
  M.~Strickland,
  Phys.\ Rev.\ Lett.\  {\bf 107}, 132301 (2011)
  [arXiv:1106.2571 [hep-ph]].
  
\bibitem{Strickland:2011aa} 
  M.~Strickland and D.~Bazow,
  Nucl.\ Phys.\ A {\bf 879}, 25 (2012)
  [arXiv:1112.2761 [nucl-th]].
  
\bibitem{mikenew}
M.~Margotta, K.~McCarty, C.~McGahan, M.~Strickland,
and D.~Yager-Elorriaga, Phys.\ Rev.\ D {\bf 83}, 105019 (2011),
arXiv:1101.4651 [hep-ph].
 


\bibitem{Mateos:2011ix} 
  D.~Mateos and D.~Trancanelli,
  Phys.\ Rev.\ Lett.\  {\bf 107}, 101601 (2011)
  [arXiv:1105.3472 [hep-th]].
  
\bibitem{Mateos:2011tv} 
  D.~Mateos and D.~Trancanelli,
  JHEP {\bf 1107}, 054 (2011)
  [arXiv:1106.1637 [hep-th]].
  
\bibitem{Gursoy:2009jd} 
  U.~Gursoy, E.~Kiritsis, L.~Mazzanti and F.~Nitti,
  Nucl.\ Phys.\ B {\bf 820}, 148 (2009)
  [arXiv:0903.2859 [hep-th]].

\bibitem{Gubser:2008ny} 
  S.~S.~Gubser and A.~Nellore,
  Phys.\ Rev.\ D {\bf 78}, 086007 (2008)
  [arXiv:0804.0434 [hep-th]].

\bibitem{Giataganas:2013lga} 
  D.~Giataganas,
  arXiv:1306.1404 [hep-th].
  
\bibitem{Fadafan:2013bva} 
  K.~B.~Fadafan, D.~Giataganas and H.~Soltanpanahi,
  arXiv:1306.2929 [hep-th].

\bibitem{AlvarezGaume:1981hn} 
  L.~Alvarez-Gaume, D.~Z.~Freedman and S.~Mukhi,
  Annals Phys.\  {\bf 134}, 85 (1981).

\bibitem{MTW}
  C.~W.~Misner, K.~S.~Thorne and J.~A.~Wheeler,
  \emph{Gravitation}.
  W.~H.~Freeman, 1973.

\bibitem{gradshteyn}
S.~Gradshteyn and I.~M.~Ryzhik; A. Jeffrey, D. Zwillinger, editors. \emph{Table of Integrals, Series, and Products}, seventh edition. Academic Press, 2007.

\bibitem{Brandhuber:1998er} 
  A.~Brandhuber, N.~Itzhaki, J.~Sonnenschein and S.~Yankielowicz,
  JHEP {\bf 9806}, 001 (1998)
  [hep-th/9803263].

\bibitem{Itzhaki:1998dd} 
  N.~Itzhaki, J.~M.~Maldacena, J.~Sonnenschein and S.~Yankielowicz,
  Phys.\ Rev.\ D {\bf 58}, 046004 (1998)
  [hep-th/9802042].
  





\end{thebibliography}
\end{document}